\documentclass[12pt]{article}
\pdfoutput=1
\usepackage{amsmath,amsfonts,graphicx,color,bbm,tikz,float,mathrsfs,amssymb,xcolor}
\usepackage{comment}
\usepackage[nosort]{cite}
\usepackage{subfigure}
\usetikzlibrary{calc,positioning}
\usetikzlibrary{patterns,arrows,decorations.pathreplacing}
\usepackage{caption}
\usepackage{ulem}
\tikzset{>=stealth}
\usepackage{lipsum}
\usepackage{tikz}

\newcommand\blfootnote[1]{%
  \begingroup
  \renewcommand\thefootnote{}\footnote{#1}%
  \addtocounter{footnote}{-1}%
  \endgroup}

\makeatletter\@addtoreset{equation}{section}\makeatother
\newcommand{\be}{\begin{equation}}
\newcommand{\ee}{\end{equation}}
\def\beq{\begin{equation}}
\def\eeq{\end{equation}}
\newcommand{\bea}{\begin{eqnarray}}
\newcommand{\eea}{\end{eqnarray}}
\newcommand{\Tr}{{\rm Tr\,}}
\newcommand{\tr}{{\rm tr\,}}

\newcommand{\vev}[1]{{\left< {#1} \right>}}
\newcommand{\bra}[1]{{\left< {#1} \right|}}
\newcommand{\ket}[1]{{\left| {#1} \right>}}

\def\nn{\nonumber}

\renewcommand{\title}[1]{\vbox{\center\LARGE{#1}}\vspace{3mm}}
\renewcommand{\author}[1]{\vbox{\center#1}\vspace{3mm}}

\newcommand{\email}[1]{\vbox{\center\tt#1}\vspace{3mm}}

\newcommand{\iti}{\tilde{i}}
\newcommand{\jti}{\tilde{j}}

\begin{document}
\begin{titlepage}
\begin{center}

{\large {\bf Local invariants of braiding quantum gates - \\ associated link polynomials and entangling power} }

\author{Pramod Padmanabhan,$^{a,\dagger}$ Fumihiko Sugino,$^b$ Diego Trancanelli$^{c,\star}$}
\blfootnote{${}^\dagger$ Current affiliation: School of Basic Sciences, Indian Institute of Technology, Bhubaneswar, India.}
\blfootnote{${}^\star$ On leave of absence from the Institute of Physics at the University of S\~ao Paulo, S\~ao Paulo, Brazil.}

\vskip -1cm
{$^a${\it Department of Physics, Sungkyunkwan University,\\ Suwon, South Korea}}
\vskip0.1cm
{ $^b${\it Center for Theoretical Physics of the Universe,\\
Institute for Basic Science, Daejeon, South Korea} 
\vskip0.1cm
$^c${\it Dipartimento di Scienze Fisiche, Informatiche e Matematiche, \\
Universit\`a di Modena e Reggio Emilia, via Campi 213/A, 41125 Modena, Italy \\ \& \\
INFN Sezione di Bologna, via Irnerio 46, 40126 Bologna, Italy}}
\email{pramod23phys, fusugino, dtrancan@gmail.com}

\vskip 1cm 
\end{center}

\abstract{
\noindent 
For a generic $n$-qubit system, local invariants under the action of $SL(2,\mathbb{C})^{\otimes n}$ characterize non-local properties of entanglement. In general, such properties are not immediately apparent and  hard to construct. Here we consider  two-qubit Yang-Baxter operators and show that their eigenvalues completely determine the non-local properties of the system. 
Moreover, we apply the Turaev procedure to these operators and obtain their associated link/knot polynomials. 
We also compute their entangling power and compare it with that of a generic two-qubit operator. 
}

\end{titlepage}
\tableofcontents 

\section{Introduction}

Entanglement, perhaps the most bizarre feature of the quantum world \cite{bell, schro}, plays a crucial role in quantum information processing and quantum computation \cite{info, qcom}. Its non-local nature goes against our classical intuition, but it can be used to analyze a quantum system in a systematic manner, via group theory and classical invariant theory \cite{weyl, taspringer}. The parameters appearing in quantum states and quantum operators can in fact be organized by their response under the local action of $SL\left(2, \mathbb{C}\right)^{\otimes n}$, for an $n$-qubit system defined on $\left(\mathbb{C}^2\right)^{\otimes n}$. 
This action defines an orbit space of equivalence classes such that the states or operators in a given orbit have the same non-local properties. This analysis has been performed in \cite{linpop1, linpop2, mmt, carteret, sudbery, 4qu, walter_etal} for local unitaries of pure and mixed states on finite-dimensional Hilbert spaces and  in \cite{yu} for two-qubit gates. As is well known from these early works, the systematic computation of these local invariants is a tedious task that gets harder as one increases the number of qubits. Nevertheless, this is important to understand all possible entanglement in a finite quantum system and hence must be tackled.

In this work we explore the possibility of simplifying this task by ``creating'' quantum systems with {\it braid operators} built from {\it Yang-Baxter operators (YBOs)}, {\it i.e.} operators that solve the (spectral parameter-independent) Yang-Baxter equation. In recent years it has been understood that such operators can also act as quantum gates  \cite{w1, w2, w3, w4, w5, w6, w7, w8, w9}, leading to the speculation of a broad connection between topological and quantum entanglement. We work on the two-qubit space ($n=2$), though we expect the properties we find to generalize to higher $n$. 

For a generic two-qubit operator, an obvious set of independent local invariants under $SL(2,\mathbb{C})^{\otimes 2}$ are class functions of the operator or functions of its independent eigenvalues, since $SL(2,\mathbb{C})^{\otimes 2}$ acts on the operator as a similarity transformation. However, this is not necessarily the whole story and there may be more independent local invariants. We expect the number of independent local invariants to reduce and get closer to the number of independent eigenvalues if some constraints on the operator are imposed.
We explicitly see that all the local invariants are solely functions of the eigenvalues for 
two-qubit braid operators of the form 
\be\label{xbraid}
\begin{pmatrix}  \;\, \bigstar \;\, & \;\, 0 \;\, & \;\, 0 \;\, & \;\, \bigstar \;\, 
\\ 0 & \bigstar & \bigstar & 0 \\ 0 & \bigstar & \bigstar & 0 \\ \bigstar & 0 & 0 & \bigstar \end{pmatrix}.
\ee
These matrices generate entangled two-qubit states and we denote them  {\it X-type operators}, for obvious reasons. We find twelve such classes of YBOs that can be both unitary and non-unitary.\footnote{In \cite{hie}, two-qubit braid operators are completely classified in ten forms, as discussed in App. \ref{app:hie}. Among these ten forms, $R_{H1,3}$ and $R_{H2,3}$ in (\ref{app:RH}) are not of the form (\ref{xbraid}). We check in App. \ref{app:rh23} that  $R_{H1,3}$ and $R_{H2,3}$ have analogous properties to the X-type YBOs.}

We organize our results as follows. In Sec.~\ref{g2qu} we find one linear and five independent quadratic invariants for an arbitrary two-qubit operator under the action of $SL(2, \mathbb{C})^{\otimes 2}$. The same procedure can also be carried out for more than two qubits. 
For the special case of an arbitrary X-type two-qubit operator, we show  in Sec.~\ref{g2quX} that independent invariants are exhausted by one linear and five quadratic invariants. 
In Sec.~\ref{X-braid} we restrict the X-type operators to braid operators and observe that all the local invariants are expressed solely as functions of the eigenvalues and that the number of independent local invariants coincides with the number of independent eigenvalues in each of  twelve possible classes. 
In Sec.~\ref{X-link} we enhance the X-type braid operators using the procedure outlined in \cite{tur} and compute their associated link/knot polynomials. 
It turns out that the polynomials are not always local invariant, although they can be expressed in terms of the eigenvalues of the braid operators.
In Sec.~\ref{ePower} we also consider the entangling powers \cite{pcl} of the X-type braid operators and compare them with the entangling power of an arbitrary X-type operator. 
We end with an outlook and discussion in Sec.~\ref{out}. In App.~\ref{app:hie} we investigate the relation between our classification of X-type YBOs and Hietarinta's classification \cite{hie}. For completeness, an analogous computation is presented  in App.~\ref{app:rh23} for families of braid operators that are not of the form (\ref{xbraid}). 

\section{$SL(2,\mathbb{C})^{\otimes 2}$ invariants of general two-qubit operators}\label{g2qu}
We consider an operator $R$ acting on two qubits $\ket{i_1\,i_2}$ as a $4\times 4$ matrix (its row and column are labeled by $(i_1\,i_2)$ and $(\iti_1\,\iti_2)$, respectively):
\be
R\ket{i_1\,i_2}=\sum_{\iti_1,\iti_2=0}^1R_{i_1\,i_2,\,\iti_1\,\iti_2}\ket{\iti_1\,\iti_2}.
\ee
When an invertible local operator (ILO) 
\be
Q=Q_1\otimes Q_2\in SL(2,\mathbb{C})^{\otimes 2}
 \label{ILO_Q}
\ee	
acts on two-qubit states ($Q R\ket{i_1\,i_2}$, $Q\ket{\iti_1\,\iti_2}$), 
we can interpret that $R$ is transformed as $QRQ^{-1}$. More precisely
\be
R_{i_1\,i_2,\,\iti_1\,\iti_2} \to \sum_{i'_1,i'_2,\iti_1^{'},\iti_2^{'}=0}^1
(Q_1)_{i_1\,i'_1}(Q_2)_{i_2\,i'_2}R_{i'_1\,i'_2,\,\iti_1^{'}\,\iti_2^{'}}\,(Q_1^{-1})_{\iti_1^{'}\,\iti_1}(Q_2^{-1})_{\iti_2^{'}\,\iti_2},
\label{changeR}
\ee
where untilded (tilded) indices with $a=1,2$, say $i_a$ ($\iti_a$), are transformed by $Q_a$ ($Q_a^{-1}$). 
From this transformation property, one can see that invariants under the action of the ILO can be constructed from a set of $R$s by contracting their indices with the four invariant tensors 
$\epsilon_{i_a\,j_a}$, $\epsilon_{\iti_a\,\jti_a}$, $\delta_{i_a\,\jti_a}$, $\delta_{\iti_a\,j_a}$, 
with  $\epsilon_{01}=-\epsilon_{10}=1$, $\epsilon_{00}=\epsilon_{11}=0$. 
Note that the resulting expressions would also be invariant under the action of a general ILO belonging to 
$\mathbb{C}^*\cdot SL(2,\mathbb{C})^{\otimes 2}$, namely Stochastic Local Operations and Classical Communication (SLOCC) \cite{dur}. 
The factor $\mathbb{C}^*$ represents the multiplication by a nonzero complex number and does not affect $R$, since these are similarity transformations. 

In the following, we present the invariants at linear and quadratic orders in $R$. Einstein's convention -- repeated indices are understood to be summed over -- is used for notational simplicity. 

\paragraph{Linear invariant}
The invariant at linear order is only one and given by
\be
I_1=R_{i_1\,i_2,\,\iti_1\,\iti_2}\delta_{i_1\,\iti_1}\delta_{i_2\iti_2}=\Tr\,R,
\ee
where $\Tr$ denotes the trace taken on the whole Hilbert space of the two qubits. 
In this case there is no possibility of contraction by $\epsilon_{i_a\,j_a}$ or $\epsilon_{\iti_a\,\jti_a}$. In order to use $\epsilon_{i_1\,j_1}$ for example, we need two untilded indices with the suffix 1, but 
$R_{i_1\, i_2,\,\iti_1\,\iti_2}$ has only one.  
 
\paragraph{Quadratic invariants}  
By exhausting all possible index contractions of $R_{i_1\,i_2,\,\iti_1\,\iti_2}\,R_{j_1\,j_2,\,\jti_1\,\jti_2}$ by the invariant tensors, we first list eight invariants which are independent of $I_1^2$:
\bea
& & I_{2,1} = R_{i_1\,i_2,\,\iti_1\,\iti_2}\,R_{j_1\,j_2,\,\jti_1\,\jti_2} \delta_{i_1\,\jti_1}\delta_{\iti_1\,j_1}\delta_{i_2\jti_2} \delta_{\iti_2\,j_2}=\Tr\,R^2, 
\\
& & I_{2,2} = R_{i_1\,i_2,\,\iti_1\,\iti_2}\,R_{j_1\,j_2,\,\jti_1\,\jti_2} \delta_{i_1\,\jti_1}\delta_{\iti_1\,j_1}\delta_{i_2\iti_2} \delta_{j_2\,\jti_2} = \tr_1\left[\left(\tr_2\,R\right)^2\right],
\\ 
& & I_{2,3} = R_{i_1\,i_2,\,\iti_1\,\iti_2}\,R_{j_1\,j_2,\,\jti_1\,\jti_2} \delta_{i_1\,\iti_1}\delta_{j_1\,\jti_1}\delta_{i_2\jti_2} \delta_{\iti_2\,j_2} = \tr_2\left[\left(\tr_1\,R\right)^2\right],
\eea
where $\tr_a$ ($a=1,2$) denotes the partial trace taken on the local Hilbert space at the $a$-th qubit;
\bea
& & I_{2,4}=R_{i_1\,i_2,\,\iti_1\,\iti_2}\,R_{j_1\,j_2,\,\jti_1\,\jti_2} \epsilon_{i_1\,j_1}\epsilon_{\iti_1\,\jti_1} \delta_{i_2\,\jti_2}\delta_{\iti_2\,j_2}
=\Tr\left[Y_1(\Theta_1R)Y_1R\right], \\
 & & I_{2,5}=R_{i_1\,i_2,\,\iti_1\,\iti_2}\,R_{j_1\,j_2,\,\jti_1\,\jti_2} \delta_{i_1\,\jti_1}\delta_{\iti_1\,j_1}\epsilon_{i_2\,j_2}\epsilon_{\iti_2\,\jti_2}
 =\Tr\left[RY_2(\Theta_2R)Y_2\right], 
\eea
where $Y_a$ is the Pauli $y$-matrix acting on the $a$-th qubit, and $\Theta_a$ represents the partial transpose with respect to indices on the $a$-th qubit;
\bea
& & I_{2,6}= R_{i_1\,i_2,\,\iti_1\,\iti_2}\,R_{j_1\,j_2,\,\jti_1\,\jti_2} \epsilon_{i_1\,j_1}\epsilon_{\iti_1\,\jti_1} \delta_{i_2\,\iti_2}\delta_{j_2\,\jti_2}
=\tr_1\left[Y_1(\tr_2\Theta_1R)Y_1(\tr_2R)\right], \\
& & I_{2,7}=R_{i_1\,i_2,\,\iti_1\,\iti_2}\,R_{j_1\,j_2,\,\jti_1\,\jti_2} \delta_{i_1\,\iti_1}\delta_{j_1\,\jti_1}\epsilon_{i_2\,j_2}\epsilon_{\iti_2\,\jti_2}
 =\tr_2\left[(\tr_1R)Y_2(\tr_1\Theta_2R)Y_2\right], \\
 & & I_{2,8}=R_{i_1\,i_2,\,\iti_1\,\iti_2}\,R_{j_1\,j_2,\,\jti_1\,\jti_2} \epsilon_{i_1\,j_1}\epsilon_{\iti_1\,\jti_1}\epsilon_{i_2\,j_2}\epsilon_{\iti_2\,\jti_2}
 =\Tr\left[R^TY_1Y_2RY_1Y_2\right].
 \eea

In addition to the eight quadratic invariants above, there are two more quadratic invariants constructed from one $R$ acting on the qubits 1 and 2 (denoted  by $R_{12}$), 
and the other $R$ acting on a qubit outside of this space, {\it e.g.} acting on the qubits 2 and 3 (labelled by the indices $j_3,\jti_3$ and denoted by $R_{23}$): 
\bea
& & I_{2,9}=R_{i_1\,i_2,\,\iti_1\,\iti_2}\,R_{j_2\,j_3,\,\jti_2\,\jti_3} \delta_{i_1\,\iti_1}\delta_{j_3\,\jti_3}\delta_{i_2\jti_2}\delta_{\iti_2\,j_2}=\tr_2\left[(\tr_1R_{12})(\tr_3R_{23})\right], \\
& & I_{2,10}=R_{i_1\,i_2,\,\iti_1\,\iti_2}\,R_{j_2\,j_3,\,\jti_2\,\jti_3} \delta_{i_1\,\iti_1}\delta_{j_3\,\jti_3}\epsilon_{i_2\,j_2}\epsilon_{\iti_2\,\jti_2}
=\tr_2\left[Y_2(\tr_1\Theta_2R_{12})Y_2(\tr_3R_{23})\right].
\eea

We should notice that $I_1^2$, $I_{2,r}$ ($r=1,\cdots,10$) are not all linearly independent. In fact, the equation 
\be
a_1I_1^2+\sum_{r=1}^{10}a_{2,r}I_{2,r}=0
\label{eq_I}
\ee
for arbitrary $R$ has the nontrivial solution
\bea
& & a_{2,6}=a_{2,1}+a_{2,2}-a_{2,4}, \qquad a_{2,7}=a_{2,1}+a_{2,3}-a_{2,5}, \qquad a_{2,8}=-a_{2,1}+a_{2,4}+a_{2,5}, \nn \\
& & a_{2,9}=a_{2,10}=-a_1-a_{2,1}-a_{2,2}-a_{2,3}.
\eea
Plugging this into (\ref{eq_I}) yields $a_1f_1(\{I\})+\sum_{s=1}^5a_{2,s}f_{2,s}(\{I\})=0$, where $f_1(\{I\})$ and $f_{2,s}(\{I\})$ ($s=1,\cdots, 5$) denote linear combinations of the quadratic invariants. 
Since this equality holds for arbitrary $a_1$ and $a_{2,s}$ ($s=1,\cdots,5$), we obtain the relations 
$f_1(\{I\})=0$ and $f_{2,s}(\{I\})=0$ ($s=1,\cdots,5$) whose explicit form is 
\bea
 &  I_1^2-I_{2,9}-I_{2,10}=0,               &  I_{2,1}+I_{2,6}+I_{2,7}-I_{2,8}-I_{2,9}-I_{2,10}=0, \nn \\
 &  I_{2,2}+I_{2,6}-I_{2,9}-I_{2,10}=0, \qquad & I_{2,3}+I_{2,7}-I_{2,9}-I_{2,10}=0, \nn \\
 &  I_{2,4}-I_{2,6}+I_{2,8}=0,              & I_{2,5}-I_{2,7}+I_{2,8}=0. 
 \label{id2}
\eea
From (\ref{id2}), we see that only five of the quadratic invariants ({\it e.g.} $I_{2,4}$, $I_{2,5}$, $I_{2,8}$, $I_{2,9}$, $I_{2,10}$) are independent.  

\section{$SL(2,\mathbb{C})^{\otimes 2}$ invariants of X-type two-qubit operators}\label{g2quX}
In this section, we consider the case that the $4\times 4$ matrix $R$ in the previous section takes the X-type form:
\be
R=\begin{pmatrix} h_1 & 0 & 0 & h_2 \\ 0 & h_3 & h_4 & 0 \\ 0 & h_5 & h_6 & 0 \\ h_7 & 0 & 0 & h_8 \end{pmatrix},
\label{RX}
\ee
which is relevant to generate entangled states. $h_i$ ($i=1,\cdots,8$) are complex parameters and the matrix eigenvalues are given by 
\be
\lambda_{1\pm}=\frac12\left[h_1+h_8\pm\sqrt{(h_1-h_8)^2+4h_2h_7}\right],\qquad 
\lambda_{2\pm}=\frac12\left[h_3+h_6\pm\sqrt{(h_3-h_6)^2+4h_4h_5}\right].
\label{EV}
\ee
Since the eigenvalues do not change under general similarity transformations for $R$, they are obviously $SL(2,\mathbb{C})^{\otimes 2}$-invariant combinations of the parameters. 

It can be seen that the invariants presented in the previous section are not simply a function of these eigenvalues, but they contain other terms, that have to be invariant combinations on their own. To check this, let us specialize the linear and quadratic invariants to (\ref{RX})
\bea
I_1 & = & h_1+h_3+h_6+h_8=\lambda_{1+}+\lambda_{1-}+\lambda_{2+}+\lambda_{2-}, \nn \\
 I_{2,4} & = & 2(h_1h_6-h_4h_5-h_2h_7+h_3h_8) \nn \\
& = &2\{\lambda_{1+}\lambda_{1-}+\lambda_{2+}\lambda_{2-}+(\lambda_{1+}+\lambda_{1-})(\lambda_{2+}+\lambda_{2-})\}  -2(h_1+h_6)(h_3+h_8), \nn \\
I_{2,5} & = &  2(h_1h_3-h_4h_5-h_2h_7+h_6h_8) \nn \\
& = & 2\{\lambda_{1+}\lambda_{1-}+\lambda_{2+}\lambda_{2-}+(\lambda_{1+}+\lambda_{1-})(\lambda_{2+}+\lambda_{2-})\}  -2(h_1+h_3)(h_6+h_8), \nn \\
I_{2,8} & = & 2(h_4h_5+h_3h_6+h_2h_7+h_1h_8) \nn \\
& = & -2\{\lambda_{1+}\lambda_{1-}+\lambda_{2+}\lambda_{2-}+(\lambda_{1+}+\lambda_{1-})(\lambda_{2+}+\lambda_{2-})\}  \nn \\
& & +2(h_1+h_3)(h_6+h_8)+2(h_1+h_6)(h_3+h_8), \nn \\
I_{2,9} & = & h_1^2+h_8^2+(h_1+h_8)(h_3+h_6)+2h_3h_6 \nn \\
& = & (h_1-h_8)^2+(h_1+h_3)(h_6+h_8)+(h_1+h_6)(h_3+h_8), \nn \\
I_{2,10} &= & h_3^2+h_6^2+(h_1+h_8)(h_3+h_6)+2h_1h_8 \nn \\
& = & (h_3-h_6)^2+(h_1+h_3)(h_6+h_8)+(h_1+h_6)(h_3+h_8), 
\label{inv_RX}
\eea
where $I_{2,1}$, $I_{2,2}$, $I_{2,3}$, $I_{2,6}$, and $I_{2,7}$ are obtained from the above through the identities (\ref{id2}). 
On the RHS in each formula of (\ref{inv_RX}), we can identify the part not expressed by the eigenvalues as an additional invariant combination.\footnote{
For example, we can show that $(h_1+h_3)(h_6+h_8)$ cannot be written in terms of the eigenvalues as follows. 
Suppose it is a function of the eigenvalues: $(h_1+h_3)(h_6+h_8)=f(\lambda_{1+},\lambda_{1-},\lambda_{2+},\lambda_{2-})$. 
Taking derivatives with respect to $h_2$ ($h_4$), we see that $f$ depends on the eigenvalues only through the combination $\lambda_{1+}+\lambda_{1-}=h_1+h_8$ 
($\lambda_{2+}+\lambda_{2-}=h_3+h_6$). Hence, $(h_1+h_3)(h_6+h_8)=f(h_1+h_8,h_3+h_6)$. 
Derivatives on the RHS with respect to $h_1$ and $h_8$ should give the same result, whereas this is not the case on the LHS. This inconsistency proves the statement. A similar proof goes for $(h_1+h_6)(h_3+h_8)$. 
}     
From $I_{2,4}$, $I_{2,5}$, $I_{2,9}$ and $I_{2,10}$, we see that $h_1-h_8$ and $h_3-h_6$ are invariants. Since $\lambda_{1+}+\lambda_{1-}=h_1+h_8$ and $\lambda_{2+}+\lambda_{2-}=h_3+h_6$ are also invariant, it is seen that $h_1$, $h_3$, $h_6$, and $h_8$ are invariant themselves. Looking at (\ref{EV}), we can conclude that the six combinations 
\be\label{linv}
h_1,\quad h_3,\quad h_6, \quad h_8,\quad h_2h_7,\quad h_4h_5
\ee
are independent $SL(2,\mathbb{C})^{\otimes 2}$ invariants. 
These can also be expressed as 
\bea
h_1 & = & \frac{1}{2}\left[\lambda_{1+} + \lambda_{1-}+ \sqrt{I_{2,9} - I_{2,8} - \frac{I_{2,4} + I_{2,5}}{2}}\right], \nonumber \\ 
h_3 & = & \frac{1}{2}\left[\lambda_{2+} + \lambda_{2-}+ \sqrt{I_{2,10} - I_{2,8} - \frac{I_{2,4} + I_{2,5}}{2}}\right], \nonumber \\
h_6 & = & \frac{1}{2}\left[\lambda_{2+} + \lambda_{2-}- \sqrt{I_{2,10} - I_{2,8} - \frac{I_{2,4} + I_{2,5}}{2}}\right], \nonumber \\
h_8 & = & \frac{1}{2}\left[\lambda_{1+} + \lambda_{1-}- \sqrt{I_{2,9} - I_{2,8} - \frac{I_{2,4} + I_{2,5}}{2}}\right], \nonumber \\
h_2h_7 & = & \frac{1}{4}\left[\left(\lambda_{1+}-\lambda_{1-}\right)^2-I_{2,9}+I_{2,8}+\frac{I_{2,4}+I_{2,5}}{2}\right], \nonumber \\
h_4h_5 & = & \frac{1}{4}\left[\left(\lambda_{2+}-\lambda_{2-}\right)^2-I_{2,10}+I_{2,8}+\frac{I_{2,4}+I_{2,5}}{2}\right], 
\eea
which are clearly functions of the eigenvalues and quadratic local invariants $I_{2,4}, I_{2,5}, I_{2,8}, I_{2,9}$ and $I_{2,10}$. 

As far as the number of independent invariants is concerned, we have two additional invariants other than the eigenvalues.   Note that, in principle, six is the lower bound of the number of the invariants, because there might appear more independent invariants when we consider invariants containing higher powers of $R$. However, by studying the dimension of orbits of the operator $R$ in  (\ref{RX})  under the action of $SL\left(2, \mathbb{C}\right)^{\otimes 2}$ as follows, one can show that the number of independent invariants is precisely six. 

The operator $R$ acting on the two qubits $i$ and $i+1$ can be expanded in terms of the Pauli-matrix basis 
as 
\begin{equation}\label{RPauli}
R = lI_iI_{i+1} + a_3~Z_i I_{i+1}+ a_6~I_iZ_{i+1} + b_9~Z_iZ_{i+1} + b_1~X_iX_{i+1}+b_2~X_iY_{i+1}+b_4~Y_iX_{i+1}+b_5~Y_iY_{i+1},
\end{equation}
where $I$ is the $2\times2$ unit matrix, and  $X$, $Y$ and $Z$ are the Pauli matrices. 
$l, a_3, a_6, b_9$ are functions of $h_1$, $h_3$, $h_6$ and $h_8$, 
whereas $b_1, b_2, b_4, b_5$ 
depend on $h_2, h_4, h_5, h_7$ as 
\begin{eqnarray}
b_1 & = & \frac{1}{4}\left(h_2 + h_4 + h_5 + h_7\right), ~~ b_2  =  \frac{\mathrm{i}}{4}\left(h_2 - h_4 + h_5 - h_7\right), \nonumber \\
b_4 & = & \frac{\mathrm{i}}{4}\left(h_2+ h_4 - h_5 - h_7\right), ~~ b_5  =  \frac{1}{4}\left(-h_2 + h_4 + h_5 - h_7\right). \label{b1-5}
\end{eqnarray}  
To study the orbits 
we consider the Lie algebra generators of $SL\left(2, \mathbb{C}\right)^{\otimes 2}$ and their commutators with $R$. 
Among six such commutators, namely $\left[X_iI_{i+1}, R\right]$, $\left[I_iX_{i+1}, R\right]$, $\left[Y_iI_{i+1}, R\right]$, $\left[I_iY_{i+1}, R\right]$, $\left[Z_iI_{i+1}, R\right]$ and $\left[I_iZ_{i+1}, R\right]$,
the first four generate terms which are not present in the original operator $R$. For example, 
\begin{eqnarray} 
\left[X_iI_{i+1}, R\right] & = & -\frac{\mathrm{i}}{2}\left(h_1+h_3-h_6-h_8\right)~Y_iI_{i+1} -\frac{\mathrm{i}}{2}\left(h_1-h_3-h_6+h_8\right)~Y_iZ_{i+1}  \nonumber \\
&  & +\frac{\mathrm{i}}{2}\left(h_2+h_4-h_5-h_7\right)~Z_iX_{i+1} +\frac{\mathrm{i}}{2}\left(-h_2+h_4+h_5-h_7\right)~Z_iY_{i+1},
\end{eqnarray}
provides terms proportional to $Y_iI_{i+1}$, $Y_iZ_{i+1}$, $Z_iX_{i+1}$ and $Z_iY_{i+1}$ which are not contained in (\ref{RPauli}). 
In what follows, we do not consider such commutators that do not preserve the X-type form.  
On the other hand, the last two commutators $\left[Z_iI_{i+1}, R\right]$ and $\left[I_iZ_{i+1}, R\right]$ preserve the X-type form and modify the coefficients $b_1$, $b_2$, $b_4$ and $b_5$ as
\begin{eqnarray}
\left[Z_iI_{i+1}, R\right] & = & b_1'~X_iX_{i+1}+b_2'~X_iY_{i+1}+b_4'~Y_iX_{i+1}+b_5'~Y_iY_{i+1},  \nn \\
\left[I_iZ_{i+1}, R\right] & = & b_1''~X_iX_{i+1}+b_2''~X_iY_{i+1}+b_4''~Y_iX_{i+1}+b_5''~Y_iY_{i+1},
\end{eqnarray}
where
\begin{eqnarray}
b_1' & = & \frac{1}{2}\left(h_2 - h_4 + h_5 - h_7\right), ~~ b_2'  =  \frac{\mathrm{i}}{2}\left(h_2 + h_4 + h_5 + h_7\right), \nonumber \\
b_4' & = & \frac{\mathrm{i}}{2}\left(h_2 - h_4 - h_5 + h_7\right), ~~ b_5'  =  \frac{1}{2}\left(-h_2 - h_4 + h_5 +h_7\right),  \label{bp1-5}
\end{eqnarray}
and
\begin{eqnarray}
b_1'' & = & \frac{1}{2}\left(h_2 +h_4 - h_5 - h_7\right), ~~ b_2''  =  \frac{\mathrm{i}}{2}\left(h_2 - h_4 - h_5 + h_7\right), \nonumber \\
b_4'' & = & \frac{\mathrm{i}}{2}\left(h_2 + h_4 + h_5 + h_7\right), ~~ b_5''  =  \frac{1}{2}\left(-h_2 + h_4 - h_5 +h_7\right). \label{b2p1-5}
\end{eqnarray}
These actions do not change the coefficients $l$, $a_3$, $a_6$ and $b_9$, consistently to $h_1$, $h_3$, $h_6$ and $h_8$ being invariants.    
Regarding the operator $R$ as a vector, the elements of the Lie algebra of $SL\left(2, \mathbb{C}\right)^{\otimes 2}$ generate six independent directions in which this vector changes, implying that the dimension of the orbit is six. 

Let us now consider only the two-dimensional orbit generated by $Z_iI_{i+1}$ and $I_iZ_{i+1}$. The orbit forms a two-dimensional surface in a four-dimensional space 
spanned by $b_1$, $b_2$, $b_4$ and $b_5$, or 
equivalently by $h_2$, $h_4$, $h_5$ and $h_7$. Since two directions perpendicular to the surface correspond to invariants under the actions, there should exist two invariant combinations made of $h_2$, $h_4$, $h_5$ and $h_7$.  
Thus, we identify the parameters $h_1, h_3, h_6, h_8$ and two combinations of the parameters $h_2$, $h_4$, $h_5$, $h_7$ 
as independent invariants, for a total of six elements. 
Note that this time six is the upper bound of the number of the independent invariants. 
These are invariants under the action of only two of the generators of 
$SL\left(2, \mathbb{C}\right)^{\otimes 2}$. When considering the action of the full generators, more constraints for the invariants may arise and the number could possibly decrease.  

Combining this with the previous assertion that six is the lower bound of the number of invariants, we conclude that there are precisely six $SL(2,\mathbb{C})^{\otimes 2}$ invariants (\ref{linv}) that one can construct out of the X-type operators. These six independent local invariants are spanned by the single linear invariant ($I_1$) and the five quadratic invariants ($I_{2,4}$, $I_{2,5}$, $I_{2,8}$, $I_{2,9}$, $I_{2,10}$) of (\ref{inv_RX}). These can also be viewed as 
``coordinates'' which label the different orbits of $R$ under the action of $SL\left(2, \mathbb{C}\right)^{\otimes 2}$.  

\section{$SL(2,\mathbb{C})^{\otimes 2}$ invariants for X-type YBOs}\label{X-braid}

In this section we consider the invariants for X-type matrices (\ref{RX}) that are YBOs, {\it i.e.} invertible solutions to the Yang-Baxter equation:
\be
(R\otimes {\bf 1}_2)({\bf 1}_2\otimes R) (R\otimes {\bf 1}_2)= ({\bf 1}_2\otimes R)(R\otimes {\bf 1}_2) ({\bf 1}_2\otimes R).
\label{ybe}
\ee
In contrast to the additional $SL(2,\mathbb{C})^{\otimes 2}$ invariants (different from the eigenvalues) found for general X-type matrices in the previous section, we find by direct inspection that for these operators here all the quadratic invariants depend only on the eigenvalues.  

We list the YBOs (except the trivial one $R\propto {\bf 1}_4$) into the following twelve classes and include the corresponding results for the quadratic invariants $I_{2,4}$, $I_{2,5}$, $I_{2,8}$, $I_{2,9}$, $I_{2,10}$ 
with the help of Mathematica: 
\begin{itemize}
\item Class 1: $h_2=h_3=h_6=h_7=0$\\
The eigenvalues are $\lambda_{1+}=h_1$, $\lambda_{1-}=h_8$, $\lambda_{2\pm}=\pm\sqrt{h_4h_5}\,(\equiv \pm \lambda_2)$ 
with the quadratic invariants
\be
I_{2,4}=I_{2,5}=-2\lambda_2^2, \qquad 
I_{2,8}=2(\lambda_{1+}\lambda_{1-}+\lambda_2^2), \qquad
I_{2,9}=\lambda_{1+}^2+\lambda_{1-}^2, \qquad 
I_{2,10}=2\lambda_{1+}\lambda_{1-}.
\ee
Note that $I_{2,8}=I_{2,10}-I_{2,4}$ and hence there are only three independent local invariants. 
This coincides with the number of independent eigenvalues: $h_1, h_8$ and $\sqrt{h_4h_5}$. 
\item Class 2: $h_1=h_4=h_5=h_8=0$, $h_6=h_3$\\
The eigenvalues are $\lambda_{1\pm}=\pm\sqrt{h_2h_7}\,(\equiv \pm\lambda_1)$, $\lambda_{2\pm}=h_3\,(\equiv \lambda_2)$ 
with the quadratic invariants 
\be
I_{2,4}=I_{2,5}=-2\lambda_1^2, \qquad I_{2,8}=2(\lambda_1^2+\lambda_2^2), \qquad I_{2,9}=I_{2,10}=2\lambda_2^2.
\ee
In this case also we have the number of independent eigenvalues, namely $h_3, \sqrt{h_2h_7}$, equal to the number of independent local invariants: $I_{2,4}, I_{2,10}$ (note in fact that $I_{2,8} = I_{2,10} - I_{2,4})$. 
\item Class 3: $h_2=h_3=0$, $h_4=-h_1$, $h_5=h_8$, $h_6=h_1+h_8$\\
The eigenvalues are $\lambda_{1+}=\lambda_{2+}=h_1\,(\equiv \lambda_+)$, $\lambda_{1-}=\lambda_{2-}=h_8\,(\equiv\lambda_-)$ 
with the quadratic invariants 
\bea
& & I_{2,4}=2\lambda_+(\lambda_++2\lambda_-),\qquad I_{2,5}=2\lambda_-(2\lambda_++\lambda_-), \qquad I_{2,8}=0, \nn \\
& & I_{2,9}=2(\lambda_+^2+\lambda_-^2+\lambda_+\lambda_-), \qquad I_{2,10}=2(\lambda_+^2+\lambda_-^2+3\lambda_+\lambda_-).
\eea
We have two independent eigenvalues in this case, $h_1$ and $h_8$, and two independent local invariants as 
\be
\lambda_+\lambda_- = \frac{I_{2,10}-I_{2,9}}{4},\qquad 2\lambda_+^2 = I_{2,4}-I_{2,10}+I_{2,9},\qquad 2\lambda_-^2 = I_{2,5}-I_{2,10}+I_{2,9},
\ee
which helps solve for $I_{2,10}-I_{2,9}$ in terms of $I_{2,4}$ and $I_{2,5}$ as
\be
I_{2,10}-I_{2,9} = \frac{2}{3}\left[I_{2,4}+I_{2,5} \pm \sqrt{(I_{2,4}+I_{2,5})^2
- 3I_{2,4}I_{2,5}}\right].
\ee

There are seven other solutions belonging to this class:
\begin{enumerate}
\item $h_2=h_3=0$, $h_4=h_1$, $h_5=-h_8$, $h_6=h_1+h_8$
\item $h_3=h_7=0$, $h_4=-h_8$, $h_5=h_1$, $h_6=h_1+h_8$
\item $h_3=h_7=0$, $h_4=h_8$, $h_5=-h_1$, $h_6=h_1+h_8$
\item $h_6=h_7=0$, $h_4=-h_1$, $h_5=h_8$, $h_3=h_1+h_8$
\item $h_6=h_7=0$, $h_4=h_1$, $h_5=-h_8$, $h_3=h_1+h_8$
\item $h_2=h_6=0$, $h_4=-h_8$, $h_5=h_1$, $h_3=h_1+h_8$
\item $h_2=h_6=0$, $h_4=h_8$, $h_5=-h_1$, $h_3=h_1+h_8$ 
\end{enumerate}
\item Class 4: $h_2=h_3=h_7=0$, $h_5=\frac{h_1}{h_4}(h_1-h_6)$, $h_8=h_1$\\
The eigenvalues are $\lambda_{1\pm}=\lambda_{2+}=h_1\,(\equiv \lambda_1)$, $\lambda_{2-}=-h_1+h_6\,(\equiv \lambda_2)$ 
with the quadratic invariants 
\bea
& & I_{2,4}=I_{2,5}=2\lambda_1(\lambda_1+2\lambda_2), \qquad I_{2,8}=2\lambda_1(\lambda_1-\lambda_2), \nn \\
& & I_{2,9}=2\lambda_1(2\lambda_1+\lambda_2),\qquad I_{2,10}=5\lambda_1^2+4\lambda_1\lambda_2+\lambda_2^2.
\eea
Here we have two independent eigenvalues, $h_1$ and $-h_1+h_6$. We can verify that only two of the four local invariants are independent by the expressions
\be 
\lambda_1^2  = \frac{I_{2,8} + I_{2,9}}{6},\qquad \lambda_2^2 = \frac{(I_{2,9}-2I_{2,8})^2}{6(I_{2,8}+I_{2,9})},
\ee
implying that $I_{2,4}$ and $I_{2,10}$ depend on $I_{2,8}$ and $I_{2,9}$. Thus we again see that the number of independent eigenvalues is the same as the number of independent local invariants.

Another solution $h_2=h_6=h_7=0$, $h_8=h_1$, $h_5=\frac{h_1}{h_4}(h_1-h_3)$ belongs to this class.  
\item Class 5: $h_2=h_3=h_7=0$, $h_5=\frac{h_1}{h_4}(h_1-h_6)$, $h_8=-h_1+h_6$\\
The eigenvalues are $\lambda_{1+}=\lambda_{2+}=h_1\,(\equiv\lambda_+)$, $\lambda_{1-}=\lambda_{2-}=-h_1+h_6\,(\equiv \lambda_-)$ 
with the quadratic invariants 
\bea
& & I_{2,4}=2\lambda_+(\lambda_++2\lambda_-), \qquad I_{2,5}=2\lambda_-(2\lambda_++\lambda_-), \qquad I_{2,8}=0, \nn \\
 & & I_{2,9}=2(\lambda_+^2+\lambda_-^2+\lambda_+\lambda_-), \qquad I_{2,10}=2(\lambda_+^2+\lambda_-^2+3\lambda_+\lambda_-).
\eea 
Once again we have two independent eigenvalues, $h_1$ and $-h_1+h_6$. We see that only two of the four local invariants are independent from the expressions,
\be
\lambda_+\lambda_- = \frac{I_{2,10}-I_{2,9}}{4}, \qquad \lambda_+^2 = \frac{I_{2,4}-I_{2,10}+I_{2,9}}{2}, \qquad \lambda_-^2 = \frac{I_{2,5}-I_{2,10}+I_{2,9}}{2},
\ee
where we can solve for $I_{2,9}-I_{2,10}$ in terms of $I_{2,4}$ and $I_{2,5}$, 
\be
I_{2,9}-I_{2,10} = \frac{-4(I_{2,4}+I_{2,5}) \pm \sqrt{16(I_{2,4}+I_{2,5})^2-48I_{2,4}I_{2,5}}}{6},
\ee
which in turn implies that the two eigenvalues, $\lambda_+$ and $\lambda_-$ are functions of $I_{2,4}$ and $I_{2,5}$. 
Thus we 
have the same number of independent local invariants and independent eigenvalues. 

Another solution $h_2=h_6=h_7=0$, $h_5=\frac{h_1}{h_4}(h_1-h_3)$, $h_8=-h_1+h_3$ belongs to this class. 
\item Class 6: $h_3=h_6=\frac{h_1+h_8}{2}$, $h_4=h_5=-\sqrt{\frac{h_1^2+h_8^2}{2}}$, $h_7=\frac{(h_1+h_8)^2}{4h_2}$\\
The eigenvalues are $\lambda_{1\pm}=\lambda_{2\pm}=\frac12\left[h_1+h_8\pm\sqrt{2(h_1^2+h_8^2)}\right]\,(\equiv\lambda_\pm)$ 
with the quadratic invariants 
\bea
& & I_{2,4}=I_{2,5}=2\lambda_+\lambda_-, \qquad I_{2,8}=2(\lambda_++\lambda_-)^2, \nn \\
 & & I_{2,9}=2(\lambda_+^2+\lambda_-^2+\lambda_+\lambda_-), \qquad I_{2,10}=2(\lambda_+^2+\lambda_-^2+3\lambda_+\lambda_-).
\eea
We can 
show that only two of these four local invariants are independent, $I_{2,4}$ and $I_{2,8}$ as can be seen from the expressions
\be
\lambda_+\lambda_- = \frac{I_{2,4}}{2},\qquad \lambda_+^2 + \lambda_-^2 = \frac{I_{2,8}}{2}-I_{2,4},
\ee
which helps solve for the independent eigenvalues, $\lambda_+$ and $\lambda_-$ in terms of $I_{2,4}$ and $I_{2,8}$. 

Another solution $h_3=h_6=\frac{h_1+h_8}{2}$, $h_4=h_5=\sqrt{\frac{h_1^2+h_8^2}{2}}$, $h_7=\frac{(h_1+h_8)^2}{4h_2}$ belongs to this class. 
\item Class 7: $h_4=h_5=-h_1$, $h_8=h_1$, $h_6=h_3$, $h_7=\frac{h_3^2}{h_2}$\\
The eigenvalues are $\lambda_{1+}=\lambda_{2+}=h_1+h_3\,(\equiv \lambda_+)$, $\lambda_{1-}=-\lambda_{2-}=h_1-h_3\,(\equiv \lambda_-)$ 
with the quadratic invariants 
\be
I_{2,4}=I_{2,5}=-2\lambda_-^2,\qquad I_{2,8}=2(\lambda_+^2+\lambda_-^2), \qquad I_{2,9}=I_{2,10}=2\lambda_+^2.
\ee
Clearly in this case we have two independent eigenvalues, $h_1+h_3$ and $h_1-h_3$ and two independent local invariants, $I_{2,4}$ and $I_{2,9}$ as $I_{2,8} = I_{2,9}-I_{2,4}$. 

Another solution $h_4=h_5=h_1$, $h_6=h_3$, $h_8=h_1$, $h_7=\frac{h_3^2}{h_2}$ belongs to this class. 
\item Class 8: $h_3=h_5=h_6=h_8=h_1$, $h_4=-h_1$, $h_7=-\frac{h_1^2}{h_2}$\\
The eigenvalues are $\lambda_{1\pm}=\lambda_{2\pm}=(1\pm \mathrm{i})h_1\,(\equiv \lambda_\pm)$ 
with the quadratic invariants 
\be
I_{2,4}=I_{2,5}=I_{2,9}=I_{2,10}=2(\lambda_++\lambda_-)^2, \qquad I_{2,8}=0.
\ee
Here we only have one local invariant which is consistent with the number of independent eigenvalues, depending on $h_1$. 

Another solution $h_3=h_4=h_6=h_8=h_1$, $h_5=-h_1$, $h_7=-\frac{h_1^2}{h_2}$ belongs to this class. 
\item Class 9: $h_2=h_3=h_6=0$, $h_8=h_1$, $h_4=h_5=-h_1$\\
The eigenvalues are $\lambda_{1\pm}=\lambda_{2+}=h_1\,(\equiv \lambda)$, $\lambda_{2-}=-h_1\,(=-\lambda)$ 
with the quadratic invariants 
\be
I_{2,4}=I_{2,5}=-2\lambda^2,\qquad I_{2,8}=4\lambda^2, \qquad I_{2,9}=I_{2,10}=2\lambda^2.
\ee
Once again we have a single local invariant and a single independent eigenvalue. 

There are three other solutions belonging to this class:
\begin{enumerate}
\item $h_3=h_6=h_7=0$, $h_8=h_1$, $h_4=h_5=-h_1$
\item $h_2=h_3=h_6=0$, $h_4=h_5=h_8=h_1$
\item $h_3=h_6=h_7=0$, $h_4=h_5=h_8=h_1$
\end{enumerate} 
%
\item Class 10: $h_2=h_3=h_6=0$, $h_4=h_5=h_8=-h_1$\\
The eigenvalues are $\lambda_{1\pm}=\lambda_{2\pm}=\pm h_1\,(\equiv \pm\lambda)$ 
with the quadratic invariants 
\be
I_{2,4}=I_{2,5}=I_{2,10}=-2\lambda^2,\qquad I_{2,8}=0, \qquad I_{2,9}=2\lambda^2.
\ee
There is a single local invariant and a single independent eigenvalue depending on $h_1$. 

There are three other solutions in this class:
\begin{enumerate}
\item $h_3=h_6=h_7=0$, $h_4=h_5=h_8=-h_1$
\item $h_2=h_3=h_6=0$, $h_4=h_5=h_1$, $h_8=-h_1$
\item $h_3=h_6=h_7=0$, $h_4=h_5=h_1$, $h_8=-h_1$
\end{enumerate}
\item Class 11: $h_2=h_6=0$, $h_1=h_5=h_8$, $h_4=-h_8$, $h_3=2h_8$\\
The eigenvalues are $\lambda_{1\pm}=\lambda_{2\pm}=h_8\,(\equiv \lambda)$ 
with the quadratic invariants 
\be
I_{2,4}=I_{2,5}=I_{2,9}=6\lambda^2,\qquad I_{2,8}=0,\qquad I_{2,10}=10\lambda^2.
\ee
The number of local invariants coincides with the number of independent eigenvalues. 

There are seven other solutions belonging to this class:
\begin{enumerate}
\item $h_6=h_7=0$, $h_1=h_5=h_8$, $h_4=-h_8$, $h_3=2h_8$
\item $h_2=h_6=0$, $h_1=h_4=h_8$, $h_5=-h_8$, $h_3=2h_8$
\item $h_6=h_7=0$, $h_1=h_4=h_8$, $h_5=-h_8$, $h_3=2h_8$
\item $h_2=h_3=0$, $h_5=h_8=h_1$, $h_4=-h_1$, $h_6=2h_1$
\item $h_3=h_7=0$, $h_5=h_8=h_1$, $h_4=-h_1$, $h_6=2h_1$
\item $h_2=h_3=0$, $h_4=h_8=h_1$, $h_5=-h_1$, $h_6=2h_1$
\item $h_3=h_7=0$, $h_4=h_8=h_1$, $h_5=-h_1$, $h_6=2h_1$
\end{enumerate}
\item Class 12: $h_4=h_5=0$, $h_3=h_6=\frac{1-\mathrm{i}}{2}h_1$, $h_8=-\mathrm{i}h_1$, $h_7=-\frac{\mathrm{i}}{2}\frac{h_1^2}{h_2}$.\\
The eigenvalues are $\lambda_{1\pm}=\lambda_{2\pm}=\frac{1-\mathrm{i}}{2}h_1\,(\equiv \lambda)$ 
with the quadratic invariants 
\be
I_{2,4}=I_{2,5}=2\lambda^2, \qquad I_{2,8}=8\lambda^2,\qquad I_{2,9}=6\lambda^2, \qquad I_{2,10}=10\lambda^2.
\ee
The number of local invariants coincides with the number of independent eigenvalues. 

Another solution $h_4=h_5=0$, $h_3=h_6=\frac{1+\mathrm{i}}{2}h_1$, $h_8=\mathrm{i}h_1$, $h_7=\frac{\mathrm{i}}{2}\frac{h_1^2}{h_2}$ belongs to this class. 
\end{itemize}

This classification is based on the pattern of eigenvalues and quadratic invariants, which is different from the criterium used by Hietarinta~\cite{hie}. The relation between the two classifications is detailed in App.~\ref{app:hie}. 

\section{Link invariants for X-type YBOs}\label{X-link}

A theorem due to Alexander \cite{alexander} states that every knot/link embedded in $S^2$ can be obtained as a closure of a braid group element. In order for this to be valid, the braid group generators must satisfy two additional moves, apart from the three usual Reidemeister moves,\footnote{Recall that the second and third Reidemeister moves represent the relations, $\sigma_i\sigma_i^{-1}=\sigma_i^{-1}\sigma_i=1$ and $\sigma_i\sigma_{i+1}\sigma_i=\sigma_{i+1}\sigma_i\sigma_{i+1}$ respectively.} called the {\it Markov moves}. This leads to the enhancement procedure of Turaev and the subsequent computation of knot/link polynomials \cite{tur}, which we perform in the following. 

\paragraph{Definition : } An {\it enhanced YBO} is a quadruple $(R, \mu, x, y)$, with $R: V\otimes V \rightarrow V\otimes V$ (a braid operator), $\mu : V \rightarrow V$ and $x, y \in \mathbb{C}^*$ such that the following conditions hold 
\begin{eqnarray}
(a)& & \left[R, \mu\otimes\mu\right]  =  0, \\
(b)&  &\tr_2\left[R~(\mu\otimes\mu)\right]   =  xy~\mu, \\
(c)&  & \tr_2\left[R^{-1}~(\mu\otimes\mu)\right]  =  x^{-1}y~\mu,
\end{eqnarray}
where, as above, $\tr_2$ denotes the partial trace on the second qubit space. 
Let $B_n$ be the $n$-strand braid group generated by $\sigma_1,\cdots,\sigma_{n-1}$.  
{\it Link polynomials} for a braid group element $\xi\in B_n$ are then obtained as 
\begin{equation}\label{lpolynomial}
L_R\left(\xi\right) = x^{-w\left(\xi\right)}y^{-n}~\Tr\left[\rho_R\left(\xi\right)\mu^{\otimes n}\right],
\end{equation}
where $w(\xi)=(\textrm{the number of positive crossings}) - (\textrm{the number of negative crossings})$ is the {\it writhe of the link}, and $\rho$ is a representation of $B_n$ constructed from the YBO $R$ as 
\begin{equation}
\rho_R(\sigma_i) = I^{\otimes i-1}\otimes R_{i,i+1}\otimes I^{\otimes n-i+1}.
\end{equation}
We take $V=\left(\mathbb{C}^2\right)^{\otimes n}$ for qubit systems and $n=2$ for two-qubit systems. 
$I={\bf 1}_2$ and $R_{i,i+1}$ denotes $R$ acting on the $i$-th and $(i+1)$-th qubits.  

Note that the polynomials obtained from (\ref{lpolynomial}) are not always invariant under the local action of $SL(2, \mathbb{C})^{\otimes n}$ due to the presence of $\mu^{\otimes n}$. 
In the case when $\mu=I$, the link polynomials are local invariants. 
As we explicitly see in (\ref{LR_ILO1}) and (\ref{LR_ILO2}), the link polynomials (\ref{lpolynomial}) with $\mu\neq I$ are not expected to be local invariants even if they are expressible only in terms of the eigenvalues. 
We can say that any local invariant constructed from an X-type YBO is expressed as a function depending only on eigenvalues of the YBO. However, the converse is not true in general.  

We  now enhance the twelve classes of X-type braid operators\footnote{
YBOs automatically become braid operators since the far-commutativity conditions $\sigma_i\sigma_j=\sigma_j\sigma_i$ ($|i-j|>1$) are trivially satisfied.} 
obtained in Sec.~\ref{X-braid} and compute the associated link invariants for two- or three-strand cases as examples. 
All of these are expressed solely in terms of the eigenvalues. Since it turns out that in each case the Skein relation for the braid operators depends only on the eigenvalues, 
we can say that all the other link invariants generated via the Skein relation are also functions only of the eigenvalues. 
Enhancement of Hietarinta's solutions \cite{hie} and associated link invariants are investigated in \cite{aizawa}. 
Cases for unitary solutions are also discussed in \cite{franko}. 
Although our results overlap with the results obtained there, we present them from our viewpoint in order to make this paper self-contained. 

\begin{itemize}
\item Class 1 : $R_1=\left(\begin{array}{cccc} h_1 & 0 & 0 & 0 \\ 0 & 0 & h_4 & 0 \\ 0 & h_5 & 0 & 0 \\ 0 & 0 & 0 & h_8 \end{array}\right)$ \\ 

In this case we can enhance the braid operator when $\mu = I$, $\mu=Z$, $\mu = I \pm Z$. 
\begin{enumerate}

\item $\mu= I$, $x=\pm h_1, ~ y=\pm 1$ and $h_8=h_1$. For example the link invariants corresponding to a two-strand braid group element $\xi=\sigma^k$ ($k\in\mathbb{Z}$) are given by
\be
L_R\left(\sigma_1^k\right) = \begin{cases} 2 + 2\left(\frac{\sqrt{h_4h_5}}{h_1}\right)^k & (\mbox{$k$ even})\\ \pm 2 & (\mbox{$k$ odd}) \end{cases}
\label{LR_class1-1}
\ee
that distinguish links with even linking numbers. 
At $h_8=h_1$ the braid operator has three eigenvalues $\left\{\lambda_1 = h_1, ~ \pm\lambda_2 = \pm \sqrt{h_4h_5}\right\}$ implying that a scaled version of this braid operator, 
$g_i =  \mp\frac{\mathrm{i}}{\sqrt{\lambda_1\lambda_2}} R$, 
realizes the Birman-Murakami-Wenzl (BMW) algebra $\mathscr{C}_n\left(l,m\right)$ \cite{bmw1, bmw2}: 
\begin{eqnarray}
&  e_i = \frac{1}{m}\left(g_i + g_i^{-1}\right)-1, \qquad &
e_i^2 = \left[\frac{1}{m}\left(l+\frac{1}{l}\right)-1\right]e_i ,\label{skein} \\
& e_ig_{i\pm 1}e_i = l e_i, \qquad & e_ig_i = g_ie_i=l^{-1}e_i,
\label{bmw2}
\end{eqnarray}
with $l=\pm\mathrm{i}\sqrt{\frac{\lambda_1}{\lambda_2}}$ and $m=\mp\mathrm{i}\frac{\lambda_1-\lambda_2}{\sqrt{\lambda_1\lambda_2}}$. 
From (\ref{skein}), we obtain 
\be
g_i^2-\left(m+\frac{1}{l}\right)g_i+\left(1+\frac{m}{l}\right)\cdot 1-\frac{1}{l}g_i^{-1}=0. 
\label{skein2}
\ee
The Skein relation for the braid operator in this case can be read off from (\ref{skein2}) as $g_i$ and $g_i^{-1}$ can be thought of as positive and negative crossings respectively. 
This helps us to obtain other link invariants in a combinatorial manner, which are also expressed in terms of the eigenvalues. 
Although the BMW algebra underlies the Kauffman polynomial of two variables \cite{kauffpol} in general, 
we see that (\ref{skein2}) with (\ref{LR_class1-1}) generates link invariants in the single variable $\frac{\sqrt{h_4h_5}}{h_1}$. 

\item $\mu =Z$,  $x=\pm h_1, ~ y=\pm 1$ and $h_8=-h_1$. The link invariants corresponding to a two-strand braid group element $\xi=\sigma_1^k$ ($k\in \mathbb{Z}$) become 
\be
L_R\left(\sigma_1^k\right) = \left[1-\left(\frac{\sqrt{h_4h_5}}{h_1}\right)^k\right]\left[1 + \left(-1\right)^k\right],
\label{L_R_class1mu=Z}
\ee
that distinguish links with even linking numbers. 
In this case, the braid operator has four eigenvalues $\left\{\pm\lambda_1 =\pm h_1, ~ \pm\lambda_2 = \pm \sqrt{h_4h_5}\right\}$ and satisfies 
the identity $R_1^3-(h_1^2+h_4h_5)R_1+h_1^2h_4h_5R_1^{-1}=0$. 
At a glance it seems to lead to nontrivial $G_2$-link invariants \cite{kup}, but actually it does not as discussed in \cite{aizawa,franko}. 

Note that (\ref{L_R_class1mu=Z}) is not a local invariant although it can be expressed in terms of the eigenvalues. Actually, under the transformation (\ref{changeR}) with 
$Q_j=\begin{pmatrix} a_ j & b_j \\ c_j & d_j \end{pmatrix}$ satisfying $a_jd_j-b_jc_j=1$ ($j=1,2$), (\ref{L_R_class1mu=Z}) changes as 
\be
2\left(\prod_{j=1}^2\left(a_jd_j+b_jc_j\right)\right)\left[1-\left(\frac{\sqrt{h_4h_5}}{h_1}\right)^k\right]
\label{LR_ILO1}
\ee
for $k$ even, and 
\be
\mp 4\frac{(h_4h_5)^{\frac{k-1}{2}}}{h_1^{k}}\left(a_1c_1b_2d_2h_4+b_1d_1a_2c_2h_5\right)
\label{LR_ILO2}
\ee
for $k$ odd. $R$ can be diagonalized as $\Omega\, {\rm diag} (\lambda_1,\lambda_2,-\lambda_2,-\lambda_1)\,\Omega^{-1}$,
where $\Omega=1\oplus \frac{1}{\sqrt{2}}\begin{pmatrix} \sqrt{h_4} & \sqrt{h_4} \\ \sqrt{h_5} & -\sqrt{h_5} \end{pmatrix} \oplus 1$ depends on $\sqrt{h_4/h_5}$ 
besides the eigenvalue $\lambda_2$. 
In general, when $\mu\neq I$, such dependence in $\Omega$ will appear in the link invariants (\ref{lpolynomial}), and the result will not be expressed only in terms of the eigenvalues. 
Here we see that $\Omega^{-1} (\mu\otimes \mu)\Omega= {\rm diag}(1,-1,-1,1)$ and that the dependence in $\Omega$ accidentally disappear. 
For this reason, the expression (\ref{L_R_class1mu=Z}) is expressed only in terms of the eigenvalues, which however does not mean the local invariance 
as seen in (\ref{LR_ILO1}) and (\ref{LR_ILO2}).   

\item\label{class1-2} $\mu = I + Z$, $x=\pm h_1, ~ y=\pm 2$, there is no relation between $h_1$ and $h_8$ in this case. The link invariants obtained in this case are just constants: 
$L_R\left(\sigma_1^k\right)=1$ for $k$ even and $\pm 1$ for $k$ odd. 
 
\item\label{class1-3} $\mu = I - Z$, $x=\pm h_8, ~ y=\pm 2$, there is no relation between $h_1$ and $h_8$ in this case. The link invariants obtained in this case are the same constants as above. 
\end{enumerate}

For \ref{class1-2} and \ref{class1-3}, the braid operators have four different eigenvalues, and the identity 
$R_1^3-(h_1+h_8)R_1^2+(h_1h_8-h_4h_5)R_1+(h_1+h_8)h_4h_5{\bf 1}-h_1h_8h_4h_5 R_1^{-1}=0$ holds. 
Despite this relation, nontrivial $G_2$-link invariants cannot be obtained. 
As discussed in \cite{aizawa}, any link invariant turns out to be $1$ or $-1$ due to the property $R_1^{\pm 1}(\mu\otimes \mu)=h_1^{\pm 1}(\mu\otimes \mu)$ for \ref{class1-2} and 
$R_1^{\pm 1}(\mu\otimes \mu)=h_8^{\pm 1}(\mu\otimes \mu)$ for \ref{class1-3}. 

\item Class 2 : $R_2=\left(\begin{array}{cccc} 0 & 0 & 0 & h_2 \\ 0 & h_3 & 0 & 0 \\ 0 & 0 & h_3 & 0 \\ h_7 & 0 & 0 & 0 \end{array}\right)$ \\ 

In this case the braid operator can be enhanced using only $\mu = I$. We then have $x=\pm h_3, ~ y =\pm 1$. The link invariants obtained are similar to the Class 1 counterpart as seen for an element of the two-strand braid group, $\xi =\sigma_1^k$ ($k\in\mathbb{Z}$):
\be
L_R\left(\sigma_1^k\right) = \begin{cases} 2 + 2\left(\frac{\sqrt{h_2h_7}}{h_3}\right)^k & (\mbox{$k$ even})\\ \pm 2 & (\mbox{$k$ odd}) ,\end{cases}
\ee
that distinguish links with even linking numbers. The braid operator has three eigenvalues $\left\{\lambda_2 = h_3, ~ \pm\lambda_1 = \pm \sqrt{h_2h_7}\right\}$ implying that a scaled version of this braid operator, $g_i = \mp \frac{\mathrm{i}}{\sqrt{\lambda_1\lambda_2}} R_2,$ realizes the BMW algebra $\mathscr{C}_n\left(l,m\right)$
at $l=\pm\mathrm{i}\sqrt{\frac{\lambda_2}{\lambda_1}}$ and $m=\mp\frac{\mathrm{i}\left(\lambda_2-\lambda_1\right)}{\sqrt{\lambda_1\lambda_2}}$.

\item Class 3 : $R_3=\left(\begin{array}{cccc} h_1 & 0 & 0 & 0 \\ 0 & 0 & -h_1 & 0 \\ 0 & h_8 & h_1 + h_8 & 0 \\ h_7 & 0 & 0 & h_8 \end{array}\right)$ \\ 

Here enhancement occurs when $\mu=Z$ and $\mu = \mp\sqrt{\frac{h_8-h_1}{h_7}} I + X-\mathrm{i}Y \mp \sqrt{\frac{h_8-h_1}{h_7}} Z$.

\begin{enumerate}

\item $\mu =Z$ and $x=\pm \mathrm{i}\sqrt{h_1h_8},~y=\mp \mathrm{i}\sqrt{\frac{h_1}{h_8}}$. The link invariants $L_R\left(\sigma_1^k\right)$ vanish in this case.

\item $\mu = -\sqrt{\frac{h_8-h_1}{h_7}} I + X-\mathrm{i}Y + \sqrt{\frac{h_8-h_1}{h_7}} Z$, and  $x=\pm h_8,~y=\mp 2\sqrt{\frac{h_8-h_1}{h_7}}$. The link invariants are constants in this case:
$L_R\left(\sigma_1^k\right)=(\pm1)^k$.

\item $\mu = \sqrt{\frac{h_8-h_1}{h_7}} I + X-\mathrm{i}Y - \sqrt{\frac{h_8-h_1}{h_7}} Z$, and  $x=\pm h_8,~y=\pm 2\sqrt{\frac{h_8-h_1}{h_7}}$. The link invariants are the same constants as above.

\end{enumerate} 

As there are two distinct eigenvalues in these cases, $\left\{h_1, h_8\right\}$, each with multiplicity two, we expect to realize the Hecke algebra, $H_n(q)$, generated by invertible $\sigma_i$, 
\be
\sigma_i^2 = \left(q-1\right)\sigma_i + q,~~\sigma_i\sigma_{i+1}\sigma_i = \sigma_{i+1}\sigma_i\sigma_{i+1},
\label{hecke}
\ee
using this braid operator \cite{jones}.
This happens either for $\sigma_i = -\frac{1}{h_1}~R_3$ at $q=-\frac{h_8}{h_1}$ or $\sigma_i = -\frac{1}{h_8}~R_3$ at $q=-\frac{h_1}{h_8}$. 
The Skein relation is read off from the first equation of (\ref{hecke}).  

\item Class 4 : $R_4=\left(\begin{array}{cccc} h_1 & 0 & 0 & 0 \\ 0 & 0 & h_4 & 0 \\ 0 & \frac{h_1}{h_4}(h_1-h_6) & h_6  & 0 \\ 0 & 0 & 0 & h_1 \end{array}\right)$ \\ 
In this case enhancement is possible when $\mu=I$, $\mu=Z$, $\mu = I \pm Z$ and when $\mu=I+\frac{h_6}{2h_1-h_6}Z$.

\begin{enumerate}

\item $\mu= I$, $x=\pm h_1,~ y=\pm 1$ and $h_6=0$. We obtain constant link invariants: 
$L_R\left(\sigma_1^k\right)= 4$ for $k$ even and $\pm 2$ for $k$ odd.  
\item $\mu=Z$, $x=\pm\mathrm{i}h_1,~y=\mp\mathrm{i}$ and $h_6=2h_1$. 
The link invariants $L_R(\sigma_1^k)$ vanish. 
\item $\mu= I + Z$, $x=\pm h_1,~ y=\pm 2$. We obtain constant link invariants $L_R\left(\sigma_1^k\right)=(\pm1)^k$.
\item $\mu= I - Z$, $x=\pm h_1,~ y=\pm 2$. We obtain the same constant link invariants as above. 

\item $\mu =  I + \frac{h_6}{2h_1-h_6}Z$ and $x=\pm \frac{h_1^{\frac{3}{2}}}{\sqrt{h_1-h_6}},~ y = \pm \frac{2\sqrt{h_1}\sqrt{h_1-h_6}}{2h_1-h_6}$. 
In this case we obtain non-trivial link polynomials as seen in a two-strand braid group element $\xi=\sigma_1^k$ ($k\in\mathbb{Z}$):
\be
L_R\left(\sigma_1^k\right) = \left[\frac{\pm h_1^{\frac{3}{2}}}{\sqrt{h_1-h_6}}\right]^{-k}~\frac{-h_1\left(-h_1+h_6\right)^{1+k}+h_1^k\left(3h_1^2-3h_1h_6+h_6^2\right)}{h_1\left(h_1-h_6\right)},
\label{LR1_class4-5}
\ee
and in a three-strand braid group element 
$\xi=\sigma_1^{k_1}\sigma_2^{k_2}$ ($k_1, k_2\in\mathbb{Z}$): 
\begin{eqnarray}
L_R\left(\sigma_1^{k_1}\sigma_2^{k_2}\right) & = &  
\left[\frac{\pm h_1^{\frac{3}{2}}}{\sqrt{h_1-h_6}}\right]^{-k_1-k_2+1}\frac{1}{h_1^3\left(2h_1^2-3h_1h_6+h_6^2\right)} \nn \\
& & \times \prod_{a=1}^2\left\{-h_1\left(-h_1+h_6\right)^{k_a+1}+h_1^{k_a}\left(3h_1^2-3h_1h_6+h_6^2\right)\right\}.
\label{LR2_class4-5}
\end{eqnarray}
\end{enumerate}

Since there are two distinct eigenvalues in these cases, $\left\{h_1, -h_1+h_6\right\}$ with multiplicities three and one respectively, 
we expect to realize the Hecke algebra (\ref{hecke}), $H_n(q)$, with this braid operator 
and this indeed happens for either $\sigma_i = \frac{1}{h_1-h_6}R_4$ at $q=\frac{h_1}{h_1-h_6}$ or $\sigma_i = -\frac{1}{h_1}R_4$ at $q=\frac{h_1-h_6}{h_1}$. 
We can show that (\ref{LR1_class4-5}) and (\ref{LR2_class4-5}) depend on the eigenvalues only through their ratio $(-h_1+h_6)/h_1$.        

\item Class 5 : $R_5=\left(\begin{array}{cccc} h_1 & 0 & 0 & 0 \\ 0 & 0 & h_4 & 0 \\ 0 & \frac{h_1}{h_4}(h_1-h_6) & h_6  & 0 \\ 0 & 0 & 0 & -h_1+h_6 \end{array}\right)$ \\ 
 
 Here enhancement occurs for 
 $\mu=Z$ and $\mu=I \pm Z$. 

 \begin{enumerate}
  \item 
$\mu=Z$, $x=\pm\sqrt{h_1(h_1-h_6)}$, $y=\pm\sqrt{\frac{h_1}{h_1-h_6}}$. 
In this case the two-strand braid group elements give vanishing link invariants. We can also see that elements of the three-strand braid group vanish:\footnote{Note that 
$L_R\left(\sigma_1^k\sigma_2^l\sigma_1^n\right)$ and  $L_R\left(\sigma_2^k\sigma_1^l\sigma_2^n\right)$ reduce to $L_R\left(\sigma_1^{k+n}\sigma_2^l\right)$ and 
$L_R\left(\sigma_1^l\sigma_2^{k+n}\right)$ respectively, because $R$ commutes with $\mu\otimes \mu$.}
$L_R\left(\sigma_1^k\sigma_2^l\right)=L_R\left(\sigma_1^k\sigma_2^l\sigma_1^m\sigma_2^n\right)=0$ for $k,l,m,n\in \mathbb{Z}$. 

\item
$\mu=I+Z$, and $x=\pm h_1$, $y=\pm 2$. 
In this case, we obtain constant link invariants: $L_R\left(\sigma_1^k\right)=(\pm 1)^k$. 
 
 \item $\mu = I - Z$, and $x=\pm (h_1-h_6)$,  
 $y=\mp 2$. 
 In this case we also obtain constant link invariants: $L_R\left(\sigma_1^k\right)=(\mp 1)^k$. 
 \end{enumerate}
For these cases, $R_5$ has two different eigenvalues $\{h_1,\,-h_1+h_6\}$ with multiplicity two for each. We see that the Hecke algebra (\ref{hecke}), $H_n(q)$, is realized by $\sigma_i=-\frac{1}{h_1}R_5$ with 
$q=\frac{h_1-h_6}{h_1}$ or $\sigma=\frac{1}{h_1-h_6}R_5$ with $q=\frac{h_1}{h_1-h_6}$. 

\item Class 6 : $R_6=\left(\begin{array}{cccc} h_1 & 0 & 0 & h_2 \\ 0 & \frac{h_1+h_8}{2} & -\sqrt{\frac{h_1^2+h_8^2}{2}} & 0 \\ 0 & -\sqrt{\frac{h_1^2+h_8^2}{2}} & \frac{h_1+h_8}{2}  & 0 \\ \frac{\left(h_1+h_8\right)^2}{4h_2} & 0 & 0 & h_8 \end{array}\right)$\\
Enhancement is possible for the following five cases ($\lambda_\pm=\frac12\left[h_1+h_8\pm\sqrt{2(h_1^2+h_8^2)}\right]$ denote the eigenvalues of $R$):
\begin{enumerate}
\item
$\mu=Z$, $x=\pm\frac{h_1-h_8}{2}$, $y=\pm 1$. 
The link invariants $L_R\left(\sigma_1^k\right)$ vanish.
\item\label{class6-2}
$\mu=I+\frac{\mathrm{i}}{2}\frac{h_1+2h_2+h_8}{\sqrt{-2h_2\lambda_-}}X+\frac12\frac{h_1-2h_2+h_8}{\sqrt{-2h_2\lambda_-}}Y-\frac{2\lambda_+}{h_1-h_8}Z$, $x=\pm \lambda_-$, $y=\pm 2$. 
\item
$\mu=I-\frac{\mathrm{i}}{2}\frac{h_1+2h_2+h_8}{\sqrt{-2h_2\lambda_-}}X-\frac12\frac{h_1-2h_2+h_8}{\sqrt{-2h_2\lambda_-}}Y-\frac{2\lambda_+}{h_1-h_8}Z$, $x=\pm \lambda_-$, $y=\pm 2$. 
\item
$\mu=I+\frac{\mathrm{i}}{2}\frac{h_1-2h_2+h_8}{\sqrt{2h_2\lambda_+}}X+\frac12\frac{h_1+2h_2+h_8}{\sqrt{2h_2\lambda_+}}Y+\frac{2\lambda_-}{h_1-h_8}Z$, $x=\pm \lambda_-$, $y=\pm 2$. 
\item\label{class6-5}
$\mu=I-\frac{\mathrm{i}}{2}\frac{h_1-2h_2+h_8}{\sqrt{2h_2\lambda_+}}X-\frac12\frac{h_1+2h_2+h_8}{\sqrt{2h_2\lambda_+}}Y+\frac{2\lambda_-}{h_1-h_8}Z$, $x=\pm \lambda_-$, $y=\pm 2$.  
\end{enumerate}
For the cases \ref{class6-2}-\ref{class6-5}, we obtain the same result for the link invariants: $L_R\left(\sigma_1^k\right)=(\pm 1)^k$. 
Each of the eigenvalues $\lambda_\pm$ has multiplicity two. 
The braid operator can be used to realize the Hecke algebra by $\sigma_1=-\frac{1}{\lambda_+}R_6$ with $q=-\frac{\lambda_-}{\lambda_+}$ or 
$\sigma_i=-\frac{1}{\lambda_-}R_6$ with $q=-\frac{\lambda_+}{\lambda_-}$. 

\item Class 7 : $R_7=\left(\begin{array}{cccc} h_1 & 0 & 0 & h_2 \\ 0 & h_3 & -h_1 & 0 \\ 0 & -h_1 & h_3  & 0 \\ \frac{h_3^2}{h_2} & 0 & 0 & h_1 \end{array}\right)$ \\  
 
The case $\mu= I$ alone enhances this operator when $x=\pm (h_1+h_3),~y=\pm 1$. We obtain non-trivial link invariants in this case as seen for two-strand and three-strand braid group elements:
\begin{eqnarray}
L_R\left(\sigma_1^k\right) & = & (\pm 1)^k \,2\left[1+\frac{1+(-1)^k}{2}\left(\frac{h_1-h_3}{h_1+h_3}\right)^k\right], \\
L_R\left(\sigma_1^k\sigma_2^l\right) & = &(\pm 1)^{k+l+1}\,2\left[1+\frac{1+(-1)^k}{2}\left(\frac{h_1-h_3}{h_1+h_3}\right)^k\right]
\left[1+\frac{1+(-1)^l}{2}\left(\frac{h_1-h_3}{h_1+h_3}\right)^l\right]. \nn \\
\end{eqnarray}
These distinguish only links with even linking numbers.\footnote{
Since the linking number is well-defined for two-component links, this statement has a meaning for $k$ even in $L_R(\sigma_1^k)$ and 
for $k/l$ even/odd or odd/even in $L_R(\sigma_1^k\sigma_2^l)$. Links obtained by taking closure of the braids have two components in the cases. 
As other cases, they have one component for $k$ odd in $L_R(\sigma_1^k)$ 
and for both $k$ and $l$ odd in $L_R(\sigma_1^k\sigma_2^l)$. Three components for both $k$ and $l$ even in $L_R(\sigma_1^k\sigma_2^l)$.}
There are three distinct eigenvalues in this case, $\left\{h_1+h_3, ~\pm(h_1-h_3)\right\}$ with multiplicities two, one and one respectively. 
The operator $g_i=\pm\frac{\mathrm{i}}{\sqrt{h_1^2-h_3^2}}R_7$ realizes the BMW algebra 
(\ref{skein}) and (\ref{bmw2}) at $l=\mp\mathrm{i}\sqrt{\frac{h_1+h_3}{h_1-h_3}}$ and $m=\pm\mathrm{i}\frac{2h_1}{\sqrt{h_1^2-h_3^2}}$. 

\item Class 8 : $R_8=\left(\begin{array}{cccc} h_1 & 0 & 0 & h_2 \\ 0 & h_1 & -h_1 & 0 \\ 0 & h_1 & h_1  & 0 \\ -\frac{h_1^2}{h_2} & 0 & 0 & h_1 \end{array}\right)$ \\  

Enhancement occurs for $\mu= I$ at $x=\pm\sqrt{2}h_1,~y=\pm\sqrt{2}$. We obtain just constant link invariants: 
$L_R\left(\sigma_1^k\right)=(\pm 1)^k\,2\cos\left(\frac{\pi}{4}k\right)$. 
There are two distinct eigenvalues, $(1\pm\mathrm{i})h_1$ leading to a realization of the Hecke algebra either when $\sigma_i = -\frac{1-\mathrm{i}}{2h_1}R_8$ at $q=\mathrm{i}$ or when $\sigma_i = -\frac{1+\mathrm{i}}{2h_1}R_8$ at $q=-\mathrm{i}$.

\item Class 9 : $R_9=\left(\begin{array}{cccc} h_1 & 0 & 0 & 0 \\ 0 & 0 & -h_1 & 0 \\ 0 & -h_1 & 0  & 0 \\ h_7 & 0 & 0 & h_1 \end{array}\right)$ \\

In this case enhancement is only possible with $\mu=I$ and $x=\pm h_1,~y=\pm 1$. The link invariants obtained are just constant: 
$L_R\left(\sigma_1^k\right)=4$ for $k$ even and $\pm 2$ for $k$ odd. There are two distinct eigenvalues in this case, $\left\{h_1, -h_1\right\}$ with multiplicities of three and one respectively. 
The Jordan decomposition of $R_9$ leads to the identity $R_9^2-h_1R_9-h_1^2{\bf 1}+h_1^3R_9^{-1}=0$ instead of the Hecke algebra realization.  

\item Class 10 : $R_{10}=\left(\begin{array}{cccc} h_1 & 0 & 0 & 0 \\ 0 & 0 & -h_1 & 0 \\ 0 & -h_1 & 0  & 0 \\ h_7 & 0 & 0 & -h_1 \end{array}\right)$ \\

We can enhance this braid operator when $\mu = Z$ and $\mu = \mp\mathrm{i}\sqrt{\frac{2h_1}{h_7}}\,I + X - \mathrm{i}Y \pm\mathrm{i}\sqrt{\frac{2h_1}{h_7}} Z$.

\begin{enumerate}

\item $\mu=Z$ and $x=\pm h_1,~y=\pm 1$. 
The link invariants $L_R\left(\sigma_1^k\right)$ vanish. 

\item $\mu = -\mathrm{i}\sqrt{\frac{2h_1}{h_7}} \,I+ X - \mathrm{i}Y +\mathrm{i}\sqrt{\frac{2h_1}{h_7}} Z$ and $x=\pm h_1,~y=\pm 2\mathrm{i}\sqrt{\frac{2h_1}{h_7}}$. 
The link invariants are constant: $L_R\left(\sigma_1^k\right)=1$ for $k$ even and $\mp 1$ for $k$ odd. 

\item $\mu = \mathrm{i}\sqrt{\frac{2h_1}{h_7}} \,I+ X - \mathrm{i}Y - \mathrm{i}\sqrt{\frac{2h_1}{h_7}} Z$ and $x=\pm h_1,~y=\mp 2\mathrm{i}\sqrt{\frac{2h_1}{h_7}}$. 
The link invariants give the same constants as above. 
\end{enumerate}

In each of these three cases the braid operator has two distinct eigenvalues, $\pm h_1$, each with multiplicity two. 
The Hecke algebra (\ref{hecke}) is realized by $\sigma_i=\pm\frac{1}{h_1}R_{10}$ with $q=1$. Then, the relation reduces to $\sigma_i^2=1$.\footnote{ 
However, $\sigma_i$ is not equivalent to the permutation operator (see (\ref{app:P}) for its matrix form), because patterns of the eigenvalues are different. 
The permutation operator has the eigenvalues $1$ and $-1$ with multiplicities three and one, respectively.}  

\item Class 11 : $R_{11}=\left(\begin{array}{cccc} h_8 & 0 & 0 & 0 \\ 0 & 2h_8 & -h_8 & 0 \\ 0 & h_8 & 0  & 0 \\ h_7 & 0 & 0 & h_8 \end{array}\right)$ \\

In this case enhancement is only possible for $\mu= Z$ at $x=\pm \mathrm{i}h_8,~y=\pm\mathrm{i}$. 
The link invariants $L_R\left(\sigma_1^k\right)$ vanish.
The braid operator has a single eigenvalue, $h_8$ with multiplicity four and the braid operator can be used to realize the Hecke algebra after scaling it with a factor $-\frac{1}{h_8}$ at $q=-1$. 
Then, the relation reduces to $(\sigma_i+1)^2=0$. 

\item Class 12 : $R_{12}=\left(\begin{array}{cccc} h_1 & 0 & 0 & h_2 \\ 0 & \frac{1-\mathrm{i}}{2}h_1 & 0 & 0 \\ 0 & 0 & \frac{1-\mathrm{i}}{2}h_1  & 0 \\ -\frac{\mathrm{i}h_1^2}{2h_2} & 0 & 0 & -\mathrm{i}h_1 \end{array}\right)$ \\

Enhancement is possible for the following five cases: 
\begin{enumerate}
\item
$\mu=Z$, $x=\pm\frac{1+\mathrm{i}}{2}h_1$, $y=\pm 1$. The link invariants $L_R\left(\sigma_1^k\right)$ vanish. 
\item\label{class12-2}
$\mu=I -\frac{h_1+(1+\mathrm{i})h_2}{\sqrt{2(1+\mathrm{i})h_1h_2}}\,X+\mathrm{i}\frac{h_1-(1+\mathrm{i})h_2}{\sqrt{2(1+\mathrm{i})h_1h_2}}\,Y+\mathrm{i}Z$, $x=\pm\frac{1-\mathrm{i}}{2}h_1$, $y=\pm 2$. 
\item 
$\mu=I +\frac{h_1+(1+\mathrm{i})h_2}{\sqrt{2(1+\mathrm{i})h_1h_2}}\,X-\mathrm{i}\frac{h_1-(1+\mathrm{i})h_2}{\sqrt{2(1+\mathrm{i})h_1h_2}}\,Y+\mathrm{i}Z$, $x=\pm\frac{1-\mathrm{i}}{2}h_1$, $y=\pm 2$. 

\item 
$\mu=I -\mathrm{i}\frac{h_1-(1+\mathrm{i})h_2}{\sqrt{2(1+\mathrm{i})h_1h_2}}\,X-\mathrm{i}\frac{h_1+(1+\mathrm{i})h_2}{\sqrt{2(1+\mathrm{i})h_1h_2}}\,Y-\mathrm{i}Z$, $x=\pm\frac{1-\mathrm{i}}{2}h_1$, $y=\pm 2$.  
\item\label{class12-5} 
$\mu=I +\mathrm{i}\frac{h_1-(1+\mathrm{i})h_2}{\sqrt{2(1+\mathrm{i})h_1h_2}}\,X+\mathrm{i}\frac{h_1+(1+\mathrm{i})h_2}{\sqrt{2(1+\mathrm{i})h_1h_2}}\,Y-\mathrm{i}Z$, $x=\pm\frac{1-\mathrm{i}}{2}h_1$, $y=\pm 2$.  
\end{enumerate}
For the cases \ref{class12-2}-\ref{class12-5}, we obtain the same result for the link invariants: $L_R\left(\sigma_1^k\right)=(\pm 1)^k$. 
The braid operator has a single eigenvalue, $\frac{1-\mathrm{i}}{2}h_1$, of multiplicity four, and can be used to realize the Hecke algebra at $q=-1$ after scaling it with a factor, $-\frac{1+\mathrm{i}}{h_1}$. 
Again, the relation reduces to $(\sigma_i+1)^2=0$. 
\end{itemize}

For all the cases, link invariants directly computed or generated via the Skein relations are functions of a single variable that is a combination of the eigenvalues of the braid operator. 
This seems to match the claim in \cite{aizawa} -- the best invariant of links obtained from the enhanced YBOs is the Jones polynomial.  

\section{Entangling power}\label{ePower}

We have seen in Sec.~\ref{X-braid} that the independent local invariants for the X-type YBOs are functions of just their independent eigenvalues, implying that in these systems the quantum entanglement and its non-local properties are obtained  in terms of the eigenvalues of the ``entanglers''. A subtle feature, as  we observed, is that this is not true for an entangler that is not a YBO. As a further check of this, 
we  compute here the {\it entangling powers} \cite{pcl} of the X-type YBOs and compare it with the entangling power of an arbitrary X-type entangler. 

The entangling power for an operator $U$ is defined as
\be \label{eP}
e_P(U) = \overline{E\left(U\ket{\psi_1}\otimes\ket{\psi_2}\right)},
\ee
where the overline denotes an average over some distribution of the product states, $\ket{\psi_1}\otimes\ket{\psi_2}$ and $E$ denotes an entanglement measure for two-qubit states. 
To determine the entanglement measure in a two-qubit space, we look for independent local invariants under the action of $SL(2,\mathbb{C})^{\otimes 2}$. 
The entanglement measure we choose to compute the entangling power is expected to be a function of only these local invariants. 

\subsection{Invariant of two-qubit states under $SL(2,\mathbb{C})^{\otimes 2}$}
 A two-qubit state 
 \be
 \ket{\psi}=\sum_{i_1,i_2=0}^1t_{i_1\,i_2}\ket{i_1\,i_2}
 \ee
 with coefficients $t_{i_1\,i_2}$ is changed by an ILO
 $
 Q=Q_1\otimes Q_2 \in SL(2,\mathbb{C})^{\otimes 2}$
 as
 \be
 Q\ket{\psi}=\sum_{i_1,i_2,j_1,j_2=0}^1 t_{i_1\,i_2}(Q_1)_{i_1\,j_1}(Q_2)_{i_2\,j_2}\ket{j_1\,j_2},
 \ee
which amounts to the change of the coefficients:
\be
t_{i_1\,i_2}\to \sum_{j_1,\,j_2=0}^1 t_{j_1\,j_2}(Q_1)_{j_1\,i_1}(Q_2)_{j_2\,i_2}.
\label{change_t}
\ee 

Invariant quantities under the change (\ref{change_t}) can be constructed by contracting indices of the coefficients by invariant tensors $\epsilon_{i_a\,j_a}$ ($a=1,2$) 
for $SL(2,\mathbb{C})$ at the $a$-th qubit. 
The invariant of the lowest order is quadratic in $t$:
\be
J_2=t_{i_1\,i_2}t_{j_1\,j_2} \epsilon_{i_1\,j_1}\epsilon_{i_2\,j_2}=2\det t.
\label{J2}
\ee

One can show that there is no independent invariant at higher orders in $t$ as follows. 
It is easy to see that we cannot construct invariants of odd orders in $t$.  
Any invariant of the $2N$-th order in $t$ can be expressed as 
\be
J_{2N}=  t_{i_1\,i_2}t_{j_1\,j_2} \epsilon_{i_1\,j_1}(K_{2N-2})_{i_2\,j_2},
\ee
where $(K_{2N-2})_{i_2\,j_2}$ denotes a polynomial of the $(2N-2)$-th order in $t$ with indices other than $i_2$ and $j_2$ contracted. 
We assume that invariants up to the order less than $2N$ are functions of $J_2$. 
Due to the identity
\be
t_{i_1\,i_2}t_{j_1\,j_2} \epsilon_{i_1\,j_1}=(\det t)\,\epsilon_{i_2\,j_2}=\frac12J_2\,\epsilon_{i_2\,j_2},
\ee
we obtain 
\be
J_{2N}=\frac12J_2\,\epsilon_{i_2\,j_2}(K_{2N-2})_{i_2\,j_2}.
\ee
Note that $\epsilon_{i_2\,j_2}(K_{2N-2})_{i_2\,j_2}$ is an invariant of the $(2N-2)$-th order and thus a function of $J_2$ by the assumption. 
Hence, $J_{2N}$ is also a function of $J_2$, which completes a proof by the induction.   

As another proof, we show that there is just a single local invariant for a two-qubit state, by considering the infinitesimal action of $SL(2, \mathbb{C})^{\otimes 2}$ on an arbitrary two-qubit state, 
as we mentioned below (\ref{RPauli}). This is obtained from the expressions 
\be
X_iI_{i+1}\left(\begin{array}{c} \alpha_1 \\ \alpha_2 \\ \alpha_3 \\ \alpha_4\end{array}\right) = \left(\begin{array}{c} \alpha_3 \\ \alpha_4 \\ \alpha_1 \\ \alpha_2\end{array}\right),\qquad 
Y_iI_{i+1}\left(\begin{array}{c} \alpha_1 \\ \alpha_2 \\ \alpha_3 \\ \alpha_4\end{array}\right) = \left(\begin{array}{c} -\mathrm{i}\alpha_3 \\ -\mathrm{i}\alpha_4 \\ \mathrm{i}\alpha_1 \\ \mathrm{i}\alpha_2\end{array}\right),\qquad  
Z_iI_{i+1}\left(\begin{array}{c} \alpha_1 \\ \alpha_2 \\ \alpha_3 \\ \alpha_4\end{array}\right) = \left(\begin{array}{c} \alpha_1 \\ \alpha_2 \\ -\alpha_3 \\ -\alpha_4\end{array}\right),
\ee
and 
\be
I_iX_{i+1}\left(\begin{array}{c} \alpha_1 \\ \alpha_2 \\ \alpha_3 \\ \alpha_4\end{array}\right) = \left(\begin{array}{c} \alpha_2 \\ \alpha_1 \\ \alpha_4 \\ \alpha_3\end{array}\right),\qquad 
I_iY_{i+1}\left(\begin{array}{c} \alpha_1 \\ \alpha_2 \\ \alpha_3 \\ \alpha_4\end{array}\right) = \left(\begin{array}{c} -\mathrm{i}\alpha_2 \\ \mathrm{i}\alpha_1 \\ -\mathrm{i}\alpha_4 \\ \mathrm{i}\alpha_3\end{array}\right),\qquad  
I_iZ_{i+1}\left(\begin{array}{c} \alpha_1 \\ \alpha_2 \\ \alpha_3 \\ \alpha_4\end{array}\right) = \left(\begin{array}{c} \alpha_1 \\ -\alpha_2 \\ \alpha_3 \\ -\alpha_4\end{array}\right),
\ee
with $i$ and $i+1$ denoting the first and second qubits respectively. It can be checked that only three of these six vectors are linearly independent. 
The three vectors generate a three-dimensional hypersurface in the four dimensions spanned by $\alpha_1,\cdots,\alpha_4$. A single direction perpendicular to the hypersurface corresponds to 
a single local invariant. 

Note that a general ILO belongs to $\mathbb{C}^*\cdot SL(2,\mathbb{C})^{\otimes 2}$ rather than $SL(2,\mathbb{C})^{\otimes 2}$. 
Due to the overall factor $\mathbb{C}^*$ (multiplication by a nonzero complex number), 
only the value of $J_2$ being zero or non-zero has an SLOCC-invariant meaning and labels SLOCC classes. 
For instance, $J_2\neq 0(=0)$ indicates the Bell-state class (the product-state class).   

\subsection{Entangling power for a general X-type two-qubit operator}
Consider a general two-qubit product state $\ket{P}=\left(a_1\ket{0}+b_1\ket{1}\right)\otimes\left(a_2\ket{0}+b_2\ket{1}\right)$ with  unit norm. 
The X-type two-qubit operator in (\ref{RX}) acts on $\ket{P}$ to give 
\begin{eqnarray}
R\ket{P} & = & \left(a_1a_2h_1 + b_1b_2h_2\right)\ket{00} + \left(a_1b_2h_3 + b_1a_2h_4\right)\ket{01} \nonumber \\ & + & \left(a_1b_2h_5 + b_1a_2h_6\right)\ket{10} + \left(a_1a_2h_7 + b_1b_2h_8\right)\ket{11}. 
\end{eqnarray}
The local invariant under $SL(2,\mathbb{C})^{\otimes 2}$ for this state is given by 
\begin{eqnarray}
\det t & = & h_1h_7a_1^2a_2^2 + h_2h_8b_1^2b_2^2 - h_3h_5a_1^2b_2^2 - h_4h_6 b_1^2a_2^2\nonumber \\
 & + & \left(h_1h_8 + h_2h_7 - h_3h_6 - h_4h_5\right)a_1a_2b_1b_2.
 \end{eqnarray}
 We choose $|\det t|^2$ as our entanglement measure,\footnote{
 In \cite{pcl}, the linear entropy $1-\tr_1\rho^2$ with $\rho$ being the reduced density matrix of a two-qubit pure state $\rho\equiv \frac{1}{\vev{\Psi |\Psi}}\tr_2\ket{\Psi}\bra{\Psi}$ is used as entanglement measure. 
 The linear entropy of $\ket{\Psi}=\sum_{i_1,i_2=0}^1t_{i_1\,i_2}\ket{i_1\,i_2}$ is computed as $2\frac{\left|\det t\right|^2}{\left[\Tr(tt^\dagger)\right]^2}$. When the state $\ket{\Psi}$ is normalized, the denominator is 1 and the expression coincides with $\left|\det t\right|^2$ up to the numerical factor 2.
 } 
and use the parametrization
 \be
a_1 = e^{\mathrm{i}\phi_1}\cos \theta_1,\qquad  b_1 = e^{-\mathrm{i}\phi_1}\sin \theta_1, 
\qquad
a_2 = e^{\mathrm{i}\phi_2}\cos \theta_2,\qquad b_2 = e^{-\mathrm{i}\phi_2}\sin \theta_2, 
\ee
which fixes the overall phase of each of the one-qubit states $a_i\ket{0}+b_i\ket{1}$ ($i=1,2$). 
Each one qubit corresponds to a point on the unit sphere (Bloch sphere) as 
\bea
r_x^{(i)} &\equiv & \left(a_i^*\bra{0}+b_i^*\bra{1}\right)X\left(a_i\ket{0}+b_i\ket{1}\right)=\sin(2\theta_i)\cos(-2\phi_i), \nn \\
r_y^{(i)} &\equiv & \left(a_i^*\bra{0}+b_i^*\bra{1}\right)Y\left(a_i\ket{0}+b_i\ket{1}\right)=\sin(2\theta_i)\sin(-2\phi_i), \nn \\
r_z^{(i)} &\equiv & \left(a_i^*\bra{0}+b_i^*\bra{1}\right)Z\left(a_i\ket{0}+b_i\ket{1}\right)=\cos(2\theta_i) ,
\eea
where we see that $(2\theta_i, -2\phi_i)$ parametrizes the unit sphere for each $i=1,2$. 
Under a uniform distribution on the Bloch spheres, namely averaging as 
\be
\overline{x}\equiv \frac{4}{\pi^2}\int_{-\pi}^0d\phi_1\int_{-\pi}^0 d\phi_2\int^{\pi/2}_0d\theta_1\sin\theta_1\cos\theta_1\int^{\pi/2}_0d\theta_2\sin\theta_2\cos\theta_2\,x(\phi_1,\phi_2,\theta_1,\theta_2),
\ee
we find the entangling power as 
 \begin{eqnarray}
 e_P(R) & = & \frac19\left[|h_1h_7|^2 + |h_2h_8|^2 + |h_3h_5|^2 + |h_4h_6|^2\right] \nonumber \\
 & & + \frac16 |h_1h_8 + h_2h_7 - h_3h_6 - h_4h_5|^2.
  \label{ePx}
 \end{eqnarray}
 Whereas the term on the second line consists only of $SL(2,\mathbb{C})^{\otimes 2}$ invariant combinations (\ref{linv}), the terms on the first line do not. 

$R$ in (\ref{RX}) becomes unitary when 
\begin{eqnarray}
& & h_1 = r_1e^{\mathrm{i}\varphi_1},  \qquad h_2 = \sqrt{1-r_1^2}\,e^{\mathrm{i}\varphi_2}, \qquad 
 h_3 = r_3e^{\mathrm{i}\varphi_3}, \qquad h_4 = \sqrt{1-r_3^2}e^{\mathrm{i}\varphi_4}, \nn \\
& & h_5 = -\sqrt{1-r_3^2}\,e^{\mathrm{i}\left(\varphi_3+\varphi_6-\varphi_4\right)},\qquad h_6 = r_3e^{\mathrm{i}\varphi_6}, \qquad 
h_7 = -\sqrt{1-r_1^2}\,e^{\mathrm{i}\left(\varphi_1+\varphi_8-\varphi_2\right)},\qquad h_8 = r_1e^{\mathrm{i}\varphi_8}, \nn \\
\end{eqnarray}
with $r_1$, $r_2$, $r_3\in[0,1]$ and $\varphi_1$, $\varphi_2$, $\varphi_3$, $\varphi_4$, $\varphi_6$, $\varphi_8\in[0,2\pi]$. 
Corresponding to (\ref{linv}), we see that 
\be
r_1,\qquad r_3,\qquad \varphi_1,\qquad \varphi_3,\qquad \varphi_6,\qquad \varphi_8
\label{linv_unitary}
\ee
are $SL(2,\mathbb{C})^{\otimes 2}$-invariant parameters. 
Then it is easy to verify that $e_P(R)$ depends only on 
such local invariants.  This is consistent with known results about the entangling power of unitary quantum gates \cite{entp1}. 

\subsection{Entangling power of X-type YBOs}

We discuss the entangling power of the twelve classes of X-type YBOs separately. 
We see that although the entangling power is not always a function only of eigenvalues for general YBOs, 
it is always so for unitary YBOs.  

\begin{itemize}

\item Class 1 : \\
The YBO $R_1$ has four free parameters, $h_1$, $h_4$, $h_5$, $h_8$, and its eigenvalues are $\lambda_{1+}=h_1$, $\lambda_{1-}=h_8$, $\pm\lambda_2=\pm\sqrt{h_4h_5}$. 
The entangling power (\ref{ePx}) reads 
\be
e_P(R_1) = \frac16\left|h_1h_8 - h_4h_5\right|^2 
=\frac16\left| \lambda_{1+}\lambda_{1-}-\lambda_2^2\right|^2,
\ee
which is a function of only the local invariants, as expected, and can be expressed only by the eigenvalues.  
The enhancement procedure possibly imposes a relation $h_1=h_8$ or $h_1=-h_8$, which however does not affect the above properties. 
The unitary YBO with $|h_1|=|h_4|=|h_5|=|h_8|=1$ also preserves the properties.  

\item Class 2 : \\
Free parameters of the YBO $R_2$ are 
$h_2$, $h_3$, $h_7$, and its eigenvalues are $\pm\lambda_1=\pm\sqrt{h_2h_7}$, $\lambda_2=h_3$. The entangling power 
\be
e_P(R_2) = \frac16\left|h_2h_7 - h_3^2\right|^2
=\frac16\left|\lambda_1^2-\lambda_2^2\right|^2,
\ee
is a function of only the local invariants,  expressed only in terms of the eigenvalues.
These properties are not changed by enhancement or by imposing the unitary condition $|h_2|=|h_3|=|h_7|=1$. 

\item Class 3 : \\
The YBO $R_3$ is a function of $h_1$, $h_7$, $h_8$, and its eigenvalues are $\lambda_+=h_1$, $\lambda_-=h_8$, which are not changed by the enhancement. 
The entangling power is computed to be
\bea\label{epr3}
 e_P(R_3)  & =  & \frac19\left[\left|h_1h_7\right|^2 + \left|h_1(h_1+h_8)\right|^2 \right] + \frac23 |h_1h_8|^2 \nn \\
& = &  \frac19\left[\left|\lambda_+h_7\right|^2+\left|\lambda_+(\lambda_++\lambda_-)\right|^2\right]+\frac23\left|\lambda_+\lambda_-\right|^2, 
\eea
 which is now dependent on $h_7$, a parameter that changes under the local action of $SL(2, \mathbb{C})^{\otimes 2}$. 
 $R_3$ is unitary for
$h_1 = -h_8=e^{\mathrm{i}\varphi_1}$ and $h_7=0$, 
turning it into a special case of Class 1. 
Then (\ref{epr3}) becomes a constant $\frac23$, which is a trivial function of the eigenvalues. 

\item Class 4 : \\
The YBO $R_4$ has three parameters $h_1$, $h_4$ and $h_6$, with its eigenvalues $\lambda_1=h_1$ and $\lambda_2=-h_1+h_6$, which is kept intact by enhancement. 
The entangling power is computed to be
\be
 e_P(R_4)  =  \frac19 |h_4h_6|^2  + \frac16 |h_1h_6|^2
 =\frac19\left|(\lambda_1+\lambda_2)h_4\right|^2+\frac16\left|\lambda_1(\lambda_1+\lambda_2)\right|^2, 
\ee
 which depends on $h_4$, a parameter that changes under the local action of $SL(2, \mathbb{C})^{\otimes 2}$. 
 $R_4$ becomes unitary when $|h_1|=|h_4|=1$ and $h_6=0$. Then the entangling power vanishes, which implies that the unitary $R_4$ is not an entangler. 

\item Class 5 : \\
The YBO $R_5$ is again a function of $h_1$, $h_4$ and $h_6$, with its eigenvalues $\lambda_+= h_1$ and $\lambda_-=-h_1+h_6$, before and after enhancement. 
The entangling power becomes 
\be
 e_P(R_5)  =  \frac19 |h_4h_6|^2  + \frac23 |h_1(h_1-h_6)|^2 
 =\frac19\left|(\lambda_++\lambda_-)h_4\right|^2+\frac23\left|\lambda_+\lambda_-\right|^2, 
\ee
which contains $h_4$, a parameter that changes under the 
$SL(2, \mathbb{C})^{\otimes 2}$. 
$R_5$ becomes unitary when $|h_1|=|h_4|=1$ and $h_6=0$,  
making it a special case of Class 1. 
Then, the entangling power becomes the constant $\frac23$.  

\item Class 6 : \\
The YBO $R_6$ has three parameters $h_1$, $h_2$, $h_8$, and its eigenvalues are given by $\lambda_\pm=\frac12\left[h_1+h_8\pm\sqrt{2(h_1^2+h_8^2)}\right]$, before and after the enhancement. 
The entangling power becomes
\be
 e_P(R_6) = \frac19\left[\left|h_2h_8\right|^2+\frac{1}{16}\left|\frac{h_1}{h_2}(h_1+h_8)^2\right|^2+\frac14\left|(h_1+h_8)\sqrt{h_1^2+h_8^2}\right|^2\right] 
 +\frac{1}{24}\left|h_1-h_8\right|^4 ,
 \ee
 which is now dependent on $h_2$, a parameter that changes under the 
 $SL(2, \mathbb{C})^{\otimes 2}$. 
 $h_1$ and $h_8$ are expressed by the eigenvalues as 
 $h_1=\frac{\lambda_++\lambda_-}{2}\pm\sqrt{-\lambda_+\lambda_-}$ and 
 $h_8=\frac{\lambda_++\lambda_-}{2}\mp\sqrt{-\lambda_+\lambda_-}$. 
 In this case $R_6$ cannot be unitary for any choice of the parameters. 
 
 \item Class 7 : \\
The YBO $R_7$ is a function of $h_1$, $h_2$ and $h_3$, with its eigenvalues $\lambda_+=h_1+ h_3$ and $\pm\lambda_-=\pm(h_1-h_3)$, which is not changed by enhancement. 
The entangling power is computed to be
\bea\label{epr7}
 e_P(R_7)   & = &  \frac19\left[|h_1h_2|^2 + \frac{|h_1h_3^2|^2}{|h_2|^2} + 2|h_1h_3|^2\right] ,
 \eea
where $h_2$ changes under the local action of $SL(2, \mathbb{C})^{\otimes 2}$. 
$h_1$ and $h_3$ are expressed by the eigenvalues: $h_1=\frac12(\lambda_++\lambda_-)$ and $h_3=\frac12(\lambda_+-\lambda_-)$.
Note that the $SL(2,\mathbb{C})^{\otimes 2}$-invariant part of the second line in (\ref{ePx}) vanishes in this case. 
 
$R_7$ is unitary when 
$
h_1 = r_1e^{\mathrm{i}\varphi_1}$, $h_2 =\sqrt{1-r_1^2} ~e^{\mathrm{i}\varphi_2}$ and $h_3 = -\mathrm{i}\sqrt{1-r_1^2}~e^{\mathrm{i}\varphi_1}$. 
Then (\ref{epr7}) is dependent on just $r_1$ that is a local invariant from (\ref{linv_unitary}), and can be written in terms of the eigenvalues. 

\item Class 8 : \\
This time $R_8$ is a function of $h_1$ and $h_2$ with its eigenvalues $(1\pm \mathrm{i})h_1$, which is preserved by enhancement. The entangling power becomes 
\be\label{epr8}
 e_P(R_8)  =  \frac19\left[|h_1h_2|^2 + \frac{|h_1^3|^2}{|h_2|^2} + 2|h_1^4|^2\right],   
 \ee
 where $h_2$ changes under the $SL(2, \mathbb{C})^{\otimes 2}$. Note that the $SL(2,\mathbb{C})^{\otimes 2}$-invariant part of the second line in (\ref{ePx}) vanishes. 
 
$R_8$ is unitary for 
$h_1 = \frac{1}{\sqrt{2}}~e^{\mathrm{i}\varphi_1}$ and $h_2 = \frac{1}{\sqrt{2}}~e^{\mathrm{i}\varphi_2}$.
Then the entangling power (\ref{epr8}) becomes a constant $\frac19$, which is a trivial function of the eigenvalues. 

\item Class 9 : \\
$R_9$ is a function of $h_1$ and $h_7$ with its eigenvalues $\pm h_1$, which is not changed by enhancement. 
The entangling power is computed to be
\be
 e_P(R_9)  =  \frac19 |h_1h_7|^2,   
 \ee
 which is now dependent on $h_7$, a parameter that changes under the local action of $SL(2, \mathbb{C})^{\otimes 2}$. 
 Again the second line in (\ref{ePx}) vanishes. 
 
$R_9$ becomes unitary when 
$
h_1 = e^{\mathrm{i}\varphi_1}$ and $h_7 = 0$, 
making it a special case of Class 1. Then the entangling power vanishes, implying that the unitary $R_9$ is not an entangler. 

\item Class 10 : \\
Again, $R_{10}$ is a function of just $h_1$ and $h_7$, with its eigenvalues $\pm h_1$, before and after enhancement. The entangling power is given by
\be
 e_P(R_{10})  =  \frac19 |h_1h_7|^2 + \frac23 |h_1|^4,   
 \ee
 where $h_7$ changes under the $SL(2, \mathbb{C})^{\otimes 2}$. 
 
$R_{10}$ is unitary for 
$
h_1 = e^{\mathrm{i}\varphi_1}$ and $h_7 = 0$, 
making it a special case of Class 1. Then $e_P(R_{10})=\frac23$, a trivial function of the eigenvalues. 

\item Class 11 : \\
$R_{11}$ has free parameters $h_7$ and $h_8$, and its eigenvalue is $h_8$, which is not affected by enhancement. The entangling power is 
\be
 e_P(R_{11})  = \frac19 |h_7h_8|^2 + \frac{10}{9} |h_8|^4,   
 \ee
 where $h_7$ changes under the $SL(2, \mathbb{C})^{\otimes 2}$. 
In this case $R_{11}$ cannot be unitary.  

\item Class 12 : \\
$R_{12}$ is a function of $h_1$ and $h_2$, with its eigenvalue $\frac{1-\mathrm{i}}{2}h_1$, before and after the enhancement. The entangling power is 
\be
 e_P(R_{12})  =  \frac19\left[|h_1h_2|^2 + \frac{|h_1|^6}{4|h_2|^2}\right] + \frac{1}{6}|h_1|^4,   
 \ee
 where $h_2$ changes under the $SL(2, \mathbb{C})^{\otimes 2}$. $R_{12}$ cannot be unitary. 

\end{itemize}

The Bell matrix (\ref{RX}) with $h_1=h_2=h_3=h_4=h_6=h_8=\frac{1}{\sqrt{2}}$ and $h_5=h_7=-\frac{1}{\sqrt{2}}$ gives the entangling power $\frac19$ that is not the largest in a two-qubit system. 
For the unitary case, Classes 1, 2, 3, 5 and 10 can give the maximum value $\frac23$. 
The Bell matrix gives the largest entanglement when it acts to $\ket{00}$, $\ket{01}$, $\ket{10}$ and $\ket{11}$. But, it does not when it acts to a general product state.  

\section{Outlook}\label{out}

Quantum gates realized using braid operators are expected to create a robust entangled state from a product state. 
The entangled states thus obtained depend on parameters forming local invariants and are insensitive to local perturbations. Such parameters should characterize non-local properties of quantum entanglement. 
This criterion can be used to exclude braid operators that do not possess this property. To achieve this, it is essential to identify the complete set of parameters of local invariants for a braiding quantum gate that would determine the quantum entanglement of these systems. 
For the twelve classes of the X-type two-qubit braid operators considered in this paper,  we found that the complete set is fixed by the independent eigenvalues of these operators. This is in marked contrast with the case of a generic two-qubit operator, whose eigenvalues alone are not sufficient to determine the entanglement measures of the system. 

One of possible future directions would be to analyze robustness of entanglement \cite{vt} for braiding quantum gates and to understand how topological properties coming from the braid contribute to the robustness of the quantum entanglement. In addition, it would be crucial to check these features for multi-qubit braid operators that can be constructed using the generalized Yang-Baxter equation \cite{r1, r2} for which several solutions have been found \cite{wk, wang2, rchen, pfd1, pfd2}. 

\subsection*{Acknowledgements}
 PP and FS are supported by the Institute for Basic Science in Korea (IBS-R024-Y1, IBS-R018-D1). DT is supported in part by the INFN grant {\it Gauge and String Theory (GAST)}.

\appendix
\section{Relation to the classification by Hietarinta}
\label{app:hie}

This rather technical appendix is devoted to a comparison between our results and the ones obtained by Hietarinta in \cite{hie}.
 
\subsection{Classification by Hietarinta}

We start by summarizing Hietarinta's classification. In \cite{hie}, all solutions to the constant algebraic Yang-Baxter equation: 
\be
R_{j_1\,j_2,\,k_1\,k_2}R_{k_1\,j_3,\,l_1\,k_3}R_{k_2\,k_3,\,l_2\,l_3}= R_{j_2\,j_3,\,k_2\,k_3} R_{j_1\,k_3,\,k_1\,l_3} R_{k_1\,k_2,\,l_1\,l_2}
\label{app:algYBE}
\ee
are presented. All the indices of $R$ take value 0 or 1. Here we represent $R$ in $4\times 4$-matrix form  as\footnote{Note that, to identify the matrix (\ref{app:matrixR}) with 
the expression in \cite{hie} (see eq. (4) there), the indices 0 and 1 here should be identified with 1 and 2 in \cite{hie}, respectively. 
Pairs of indices 01 and 10 are swapped in \cite{hie}.}  
\be
R=\begin{pmatrix} R_{00,\,00} & R_{00,\,01} & R_{00,\,10} & R_{00,\,11} \\
R_{01,\,00} & R_{01,\,01} & R_{01,\,10} & R_{01,\,11} \\
R_{10,\,00} & R_{10,\,01} & R_{10,\,10} & R_{10,\,11} \\
R_{11,\,00} & R_{11,01} & R_{11,\,10} & R_{11,\,11} \end{pmatrix}.
\label{app:matrixR}
\ee
Via the replacement $R\to PR$ with $P$ being the permutation matrix 
\be
P=\begin{pmatrix} 1 & 0 & 0 & 0 \\ 0 & 0 & 1 & 0 \\ 0 & 1 & 0 & 0 \\ 0 & 0 & 0 & 1\end{pmatrix}, 
\label{app:P}
\ee
(\ref{app:algYBE}) is transcribed as the braided Yang-Baxter equation:
\be
R_{i_1\,i_2,\,j_1\,j_2}R_{j_2\,i_3,\,k_2\,l_3}R_{j_1\,k_2,\,l_1\,l_2}=R_{i_2\,i_3,\,j_2\,j_3}R_{i_1\,j_2,\,l_1\,k_2}R_{k_2\,j_3,\,l_2\,l_3}
\label{app:bYBE}
\ee
that is identical to (\ref{ybe}). 

Relevant results in \cite{hie} are summarized for solutions to (\ref{app:bYBE}) as follows. The continuous transformations 
\be
R \to \kappa (Q\otimes Q)R(Q\otimes Q)^{-1}\,,
\label{app:Rchange}
\ee
with $\kappa$ a complex factor and $Q$ an invertible $2\times 2$ matrix, map a solution to a solution. 
Each of the following discrete transformations
\bea
& & R_{i\,j,\,k\,l}\to R_{k\,l,\,i\,j} \,,  \label{app:3a} \\
& & R_{i\,j,\,k\,l} \to R_{\bar{i}\,\bar{j},\,\bar{k}\,\bar{l}} \,, \label{app:3b} \\
& & R_{i\,j,\,k\,l} \to R_{j\,i,\,l\,k} \label{app:3c}
\eea
also maps a solution to a solution, where (\ref{app:3a}) means the matrix transpose taken in (\ref{app:matrixR}), and $\bar{i}$ is the negation of $i$, i.e., 
$\bar{0}\equiv 1$ and $\bar{1}\equiv 0$ in (\ref{app:3b}). 
Up to the transformations (\ref{app:Rchange})-(\ref{app:3c}), all the invertible solutions to (\ref{app:bYBE}), except the trivial solution $R\propto {\bf 1}$, are classified by the 
ten matrices:
\bea
& & R_{H3,1} = \begin{pmatrix} k & 0 & 0 & 0 \\ 0 & 0 & p & 0 \\ 0 & q & 0 & 0 \\ 0& 0 & 0 & s \end{pmatrix} ,
\qquad 
R_{H2,1}=\begin{pmatrix} k^2 & 0 & 0 & 0 \\ 0 & k^2-pq & kp & 0 \\ 0 & kq & 0 & 0 \\ 0 & 0 & 0 & k^2 \end{pmatrix} , 
\qquad
R_{H2,2}=\begin{pmatrix} k^2 & 0 & 0 & 0 \\ 0 & k^2-pq & kp & 0 \\ 0 & kq & 0 & 0 \\ 0 & 0 & 0 & -pq \end{pmatrix} , 
\nn\\
& & R_{H2,3}=\begin{pmatrix} k & p & q & s \\ 0 & 0 & k & p \\ 0 & k & 0 & q\\ 0 & 0 & 0 & k \end{pmatrix}, 
\qquad 
R_{H1,1}=\begin{pmatrix} p^2+2pq-q^2 & 0 & 0 & p^2-q^2 \\ 0 & p^2-q^2 & p^2+q^2 & 0 \\ 0 & p^2+q^2 & p^2 -q^2 & 0 \\ p^2-q^2 & 0 & 0 & p^2-2pq-q^2 \end{pmatrix},
\nn\\
& & R_{H1,2}=\begin{pmatrix} p & 0 & 0 & k \\ 0 & p-q & p & 0 \\ 0 & q & 0 & 0 \\ 0 & 0 & 0 & -q \end{pmatrix}, 
\qquad
R_{H1,3} =\begin{pmatrix} k^2 & -kp & kp & pq \\ 0 & 0 & k^2 & kq \\ 0 & k^2 & 0 & -kq \\ 0 & 0 & 0 & k^2 \end{pmatrix},
\qquad 
R_{H1,4}=\begin{pmatrix} 0 & 0 & 0 & p \\ 0 & k & 0 & 0 \\ 0 & 0 & k & 0 \\ q & 0 & 0 & 0 \end{pmatrix},
\nn \\
& & R_{H0,1}=\begin{pmatrix} 1 & 0 & 0 & 1 \\ 0 & 0 & -1 & 0 \\ 0 & -1 & 0 & 0 \\ 0 & 0 & 0 & 1 \end{pmatrix}, 
\qquad
R_{H0,2} = \begin{pmatrix} 1 & 0 & 0 & 1 \\ 0 & 1 & 1 & 0 \\ 0 & -1 & 1 & 0 \\ -1 & 0 & 0 & 1 \end{pmatrix}. 
\label{app:RH}
\eea
For X-type solutions that we consider in the text, the classification is valid with removing $R_{H1,3}$ and setting $p=q=0$ in $R_{H2,3}$: 
\be
R_{H2,3}'=\begin{pmatrix} k & 0 & 0 & s \\ 0 & 0 & k & 0 \\ 0 & k & 0 & 0\\ 0 & 0 & 0 & k \end{pmatrix}.
\ee

\subsection{Our solutions}

Our solutions in Classes 1-12 presented in Sec. \ref{X-braid} are classified by eigenvalues and quadratic invariants. Here we classify all the nontrivial solutions in Sec. \ref{X-braid} to the nine families ($R_{H3,1}$, $R_{H2,1}$, $R_{H2,2}$, $R'_{H2,3}$, $R_{H1,1}$, $R_{H1,2}$, $R_{H1,4}$, $R_{H0,1}$, $R_{H0,2}$).  

\begin{itemize}
\item
Class 1: 
$
R_1=\begin{pmatrix} h_1 & 0 & 0 & 0 \\ 0 & 0 & h_4 & 0 \\ 0 & h_5 & 0 & 0 \\ 0 & 0 & 0 & h_8 \end{pmatrix}
$\\
This falls into $R_{H3,1}$ with $k=h_1$, $p=h_4$, $q=h_5$ and $ s=h_8$. 
\item
Class 2:
$
R_2=\begin{pmatrix} 0 & 0 & 0 & h_2 \\ 0 & h_3 & 0 & 0 \\ 0 & 0 & h_3 & 0 \\ h_7 & 0 & 0 & 0 \end{pmatrix}
$\\
This falls into $R_{H1,4}$ with $k=h_3$, $p=h_2$ and $q=h_7$. 
\item
Class 3: 
$
R_3=\begin{pmatrix} h_1 & 0 & 0 & 0 \\ 0 & 0 & -h_1 & 0 \\ 0 & h_8 & h_1+h_8 & 0 \\ h_7 & 0 & 0 & h_8 \end{pmatrix}
$\\
This falls into $R_{H1,2}$ with $k=h_7$, $p=h_8$ and $q=-h_1$ by the transformation (\ref{app:3b}).  \\
The seven other solutions given in the text are equivalent to the representative as 
\begin{enumerate}
\item
$R_{\text{3-1}}=\begin{pmatrix} h_1 & 0 & 0 & 0 \\ 0 & 0 & h_1 & 0 \\ 0 & -h_8 & h_1+h_8 & 0 \\ h_7 & 0 & 0 & h_8 \end{pmatrix}$ becomes $R_3$ after the transformations (\ref{app:3b}), (\ref{app:3c}) and (\ref{app:3a}) with the redefinition $h_1\leftrightarrow h_8$.  
\item
$R_{\text{3-2}}=\begin{pmatrix} h_1 & 0 & 0 & h_2 \\ 0 & 0 & -h_8 & 0 \\ 0 & h_1 & h_1+h_8 & 0 \\ 0 & 0 & 0 & h_8 \end{pmatrix}$ becomes $R_3$ after the transformations (\ref{app:3b}) and (\ref{app:3c}) with the redefinition $h_1\leftrightarrow h_8$ and $ h_2 \to h_7$.
\item
$R_{\text{3-3}}=\begin{pmatrix} h_1 & 0 & 0 & h_2 \\ 0 & 0 & h_8 & 0 \\ 0 & -h_1 & h_1+h_8 & 0 \\ 0 & 0 & 0 & h_8 \end{pmatrix}$ becomes $R_3$ after (\ref{app:3a}) with $h_2\to h_7$. 
\item
$R_{\text{3-4}}=\begin{pmatrix} h_1 & 0 & 0 & h_2 \\ 0 & h_1+h_8 & -h_1 & 0 \\ 0 & h_8 & h_1+h_8 & 0 \\ 0 & 0 & 0 & h_8 \end{pmatrix}$ becomes $R_{\text{3-3}}$ after (\ref{app:3b}). 
\item
$R_{\text{3-5}}=\begin{pmatrix} h_1 & 0 & 0 & h_2 \\ 0 & h_1+h_8& h_1 & 0 \\ 0 & -h_8 & 0 & 0 \\ 0 & 0 & 0 & h_8 \end{pmatrix}$ becomes $R_{\text{3-1}}$ after (\ref{app:3a}) and (\ref{app:3c}) with $h_2 \to h_7$.
\item 
$R_{\text{3-6}}=\begin{pmatrix} h_1 & 0 & 0 & 0 \\ 0 & h_1+h_8 & -h_8 & 0 \\ 0 & h_1 & 0 & 0 \\ h_7 & 0 & 0 & h_8 \end{pmatrix}$ becomes $R_{\text{3-1}}$ after (\ref{app:3c}). 
\item
$R_{\text{3-7}}=\begin{pmatrix} h_1 & 0 & 0 & 0 \\ 0 & h_1+h_8 & h_8 & 0 \\ 0 & -h_1 & 0 & 0 \\ h_7 & 0 & 0 & h_8 \end{pmatrix}$ becomes $R_3$ after (\ref{app:3c}).
\end{enumerate}
\item
Class 4: 
$
R_4=\begin{pmatrix} h_1 & 0 & 0 & 0 \\ 0 & 0& h_4 & 0 \\ 0 & \frac{h_1}{h_4}(h_1-h_6) & h_6 & 0 \\ 0 & 0 & 0 & h_1 \end{pmatrix}
$\\
This falls into $R_{H2,1}$ with $k=\sqrt{h_1}$, $p=\frac{\sqrt{h_1}}{h_4}(h_1-h_6)$ and $q=\frac{h_4}{\sqrt{h_1}}$ by the transformation (\ref{app:3c}). \\
The other solution 
$R_{\text{4-1}}=\begin{pmatrix} h_1 & 0 & 0 & 0 \\ 0 & h_3 & h_4 & 0 \\ 0 & \frac{h_1}{h_4}(h_1-h_6) & 0 & 0 \\ 0 & 0 & 0 & h_1 \end{pmatrix}$ becomes $R_4$ after (\ref{app:3a}) and (\ref{app:3c}) with $h_3\to h_6$. 
\item
Class 5:
$R_5=\begin{pmatrix} h_1 & 0 & 0 & 0 \\ 0 & 0 & h_4 & 0 \\ 0 & \frac{h_1}{h_4}(h_1-h_6) & h_6 & 0 \\ 0 & 0 & 0 & -h_1+h_6 \end{pmatrix}$\\
This falls into $R_{H2,2}$ with $k=\sqrt{h_1}$, $p=\frac{\sqrt{h_1}}{h_4}(h_1-h_6)$ and $q=\frac{h_4}{\sqrt{h_1}}$ by the transformation (\ref{app:3c}). \\
The other solution 
$R_{\text{5-1}}=\begin{pmatrix} h_1 & 0 & 0 & 0 \\ 0 & h_3 & h_4 & 0 \\ 0 & \frac{h_1}{h_4}(h_1-h_6) & 0 & 0 \\ 0 & 0 & 0 & -h_1+h_3 \end{pmatrix}$ becomes $R_5$ after (\ref{app:3a}) and (\ref{app:3c}) with $h_3\to h_6$. 
\item
Class 6:
$R_6=\begin{pmatrix} h_1 & 0 & 0 & h_2 \\ 0 & \frac{h_1+h_8}{2} & -\sqrt{\frac{h_1^2+h_8^2}{2}} & 0 \\ 0 & -\sqrt{\frac{h_1^2+h_8^2}{2}} &\frac{h_1+h_8}{2} & 0 \\ \frac{(h_1+h_8)^2}{4h_2} & 0 & 0 & h_8 \end{pmatrix}$\\
This falls into $R_{H1,1}$ with $p=\frac{h_1+h_8}{2}$ and $q=-\frac12\left[h_1+h_8+\sqrt{2(h_1^2+h_8^2)}\right]=-\lambda_+$ by the transformation (\ref{app:Rchange}). 
It can be seen that $\kappa (Q\otimes Q) R_{H1,1} (Q\otimes Q)^{-1}=R_6$ with $\kappa=-\frac{1}{2\lambda_+}$ and $Q=\begin{pmatrix} \sqrt{2h_2} & 0 \\ 0 & \sqrt{h_1+h_8} \end{pmatrix}$. \\
The other solution 
$R_{\text{6-1}}=\begin{pmatrix} h_1 & 0 & 0 & h_2 \\ 0 & \frac{h_1+h_8}{2} & \sqrt{\frac{h_1^2+h_8^2}{2}} & 0 \\ 0 & \sqrt{\frac{h_1^2+h_8^2}{2}} &\frac{h_1+h_8}{2} & 0 \\ \frac{(h_1+h_8)^2}{4h_2} & 0 & 0 & h_8 \end{pmatrix}$ becomes $-R_6$ with the redefinition $h_a\to -h_a$ ($a=1,2,8$).  
\item
Class 7:
$R_7=\begin{pmatrix} h_1 & 0 & 0 & h_2 \\ 0 & h_3 & -h_1 & 0 \\ 0 & -h_1 & h_3 & 0 \\ \frac{h_3^2}{h_2} & 0 & 0 & h_1 \end{pmatrix}$\\
This falls into $R_{H1,4}$ with $k=h_1+h_3$ and $p=q=h_1-h_3$ by the transformation (\ref{app:Rchange}): 
$\kappa (Q\otimes Q) R_{H1,4} (Q\otimes Q)^{-1}=R_7$ with 
$\kappa=1$ and $Q=\begin{pmatrix} \mathrm{i}\sqrt{h_2} & -\mathrm{i}\sqrt{h_2} \\  \sqrt{h_3} & \sqrt{h_3} \end{pmatrix}$.  \\
The other solution 
$R_{\text{7-1}}=\begin{pmatrix} h_1 & 0 & 0 & h_2 \\ 0 & h_3 & h_1 & 0 \\ 0 & h_1 & h_3 & 0 \\ \frac{h_3^2}{h_2} & 0 & 0 & h_1 \end{pmatrix}$ is not equivalent to $R_7$. 
Actually, we can see that $R_{\text{7-1}}$ falls into $R_{H3,1}$ with $p=q=h_1-h_3$ and $k=s=h_1+h_3$ by 
$\kappa (Q\otimes Q) R_{H3,1} (Q\otimes Q)^{-1}=R_{\text{7-1}}$ with $\kappa=1$ and $Q=\begin{pmatrix} \sqrt{h_2} & -\sqrt{h_2} \\  \sqrt{h_3} & \sqrt{h_3} \end{pmatrix}$. 
However, $R_7$ and $R_{\text{7-1}}$ belong to the same class in our classification, since they have the same eigenvalues and quadratic invariants. We explicitly see that they are 
$SL(2,\mathbb{C})^{\otimes 2}$-equivalent: $(Q_1\otimes Q_2) R_7 (Q_1\otimes Q_2)^{-1}=R_{\text{7-1}}$ with 
\be
Q_1=\begin{pmatrix} 1 & 0 \\ 0 & -\mathrm{i} \end{pmatrix}, 
\qquad
Q_2=  \begin{pmatrix} 1 & 0 \\ 0 & \mathrm{i} \end{pmatrix}.
\label{app:Q1Q2}
\ee
\item 
Class 8:
$R_8=\begin{pmatrix} h_1 & 0 & 0 & h_2 \\ 0 & h_1 & -h_1 & 0 \\ 0 & h_1 & h_1 & 0 \\ -\frac{h_1^2}{h_2}& 0 & 0 & h_1 \end{pmatrix}$\\
This falls into $R_{H0,2}$ by the transformation (\ref{app:Rchange}): 
$\kappa (Q\otimes Q) R_{H0,2} (Q\otimes Q)^{-1}=R_8$ 
with $\kappa=h_1$ and $Q=\begin{pmatrix} 0 & \mathrm{i}\sqrt{h_2} \\ \sqrt{h_1} & 0 \end{pmatrix}$. \\
The other solution 
$R_{\text{8-1}}=\begin{pmatrix} h_1 & 0 & 0 & h_2 \\ 0 & h_1 & h_1 & 0 \\ 0 & -h_1 & h_1 & 0 \\ -\frac{h_1^2}{h_2}& 0 & 0 & h_1 \end{pmatrix}$ becomes $R_8$ by (\ref{app:3c}). 
\item 
Class 9: 
$R_9=\begin{pmatrix} h_1 & 0 & 0 & 0 \\ 0 & 0 & -h_1 & 0 \\ 0 & -h_1 & 0 & 0 \\ h_7 & 0 & 0 & h_1 \end{pmatrix}$\\
This falls into $R_{H0,1}$ by the successive transformations (\ref{app:Rchange}) and (\ref{app:3a}): 
$\kappa (Q\otimes Q) R_{H0,1} (Q\otimes Q)^{-1}$ with 
$\kappa=h_1$ and $Q=\begin{pmatrix} \sqrt{h_7} & 0 \\ 0 & \sqrt{h_1} \end{pmatrix}$ followed by (\ref{app:3a}) gives $R_9$. \\
Among the other three solutions 
\[
R_{\text{9-1}}=\begin{pmatrix} h_1 & 0 & 0 & h_2 \\ 0 & 0 & -h_1 & 0 \\ 0 & -h_1 & 0 & 0 \\ 0 & 0 & 0 & h_1 \end{pmatrix}, \qquad 
R_{\text{9-2}}= \begin{pmatrix} h_1 & 0 & 0 & 0 \\ 0 & 0 & h_1 & 0 \\ 0 & h_1 & 0 & 0 \\ h_7 & 0 & 0 & h_1 \end{pmatrix}, \qquad
R_{\text{9-3}}=\begin{pmatrix} h_1 & 0 & 0 & h_2 \\ 0 & 0 & h_1 & 0 \\ 0 & h_1 & 0 & 0 \\ 0 & 0 & 0 & h_1 \end{pmatrix},
\]
$R_{\text{9-1}}$ becomes $R_9$ by (\ref{app:3a}) with $h_2\to h_7$, whereas $R_{\text{9-2}}$ and $R_{\text{9-3}}$ are not equivalent to the representative $R_9$. 
Actually, $R_{\text{9-2}}$ falls into $R_{H2,3}'$ with $k=h_1$ and $s=h_7$ by the transformation (\ref{app:3a}), 
and $R_{\text{9-3}}$ becomes $R_{\text{9-2}}$ by (\ref{app:3a}) with $h_2\to h_7$. 
However, these two groups are $SL(2,\mathbb{C})^{\otimes 2}$ equivalent: $(Q_1\otimes Q_2) R_9 (Q_1\otimes Q_2)^{-1}=R_{\text{9-2}}$ with (\ref{app:Q1Q2}).  
\item
Class 10: 
$R_{10}=\begin{pmatrix} h_1 & 0 & 0 & 0 \\ 0 & 0 & -h_1 & 0 \\ 0 & -h_1 & 0 & 0 \\ h_7 & 0 & 0 & -h_1 \end{pmatrix}$\\
This falls into $R_{H1,2}$ with $k=h_7$ and $p=q=-h_1$ by the transformation (\ref{app:3b}). \\
The other three solutions are equivalent to the representative $R_{10}$ as
\begin{enumerate}
\item
$R_{\text{10-1}}=\begin{pmatrix} h_1 & 0 & 0 & h_2 \\ 0 & 0 & -h_1 & 0 \\ 0 & -h_1 & 0 & 0 \\ 0 & 0 & 0 & -h_1 \end{pmatrix}$ becomes $R_{10}$ by (\ref{app:3a}) with $h_2\to h_7$.
\item
$R_{\text{10-2}}=\begin{pmatrix} h_1 & 0 & 0 & 0 \\ 0 & 0 & h_1 & 0 \\ 0 & h_1 & 0 & 0 \\ h_7 & 0 & 0 & -h_1 \end{pmatrix}$ becomes $R_{10}$ by (\ref{app:3b}) and (\ref{app:3a}) with 
$h_1\to -h_1$.
\item
$R_{\text{10-3}}=\begin{pmatrix} h_1 & 0 & 0 & h_2 \\ 0 & 0 & h_1 & 0 \\ 0 & h_1 & 0 & 0 \\ 0 & 0 & 0 & -h_1 \end{pmatrix}$ becomes $R_{\text{10-2}}$ by (\ref{app:3a}) with $h_2\to h_7$.
\end{enumerate} 
\item
Class 11: 
$R_{11}=\begin{pmatrix} h_8 & 0 & 0 & 0 \\ 0 & 2h_8 & -h_8 & 0 \\ 0 & h_8 & 0 & 0 \\ h_7 & 0 & 0 & h_8 \end{pmatrix}$\\
This falls into $R_{H1,2}$ with $k=h_7$, $p=h_8$ and $q=-h_8$ by the transformation (\ref{app:3a}). \\
The other seven solutions are equivalent to the representative $R_{11}$ as
\begin{enumerate}
\item
$R_{\text{11-1}}=\begin{pmatrix} h_8 & 0 & 0 & h_2 \\ 0 & 2h_8 & -h_8 & 0 \\ 0 & h_8 & 0 & 0 \\ 0 & 0 & 0 & h_8 \end{pmatrix}$ becomes $R_{11}$ by (\ref{app:3b}) and (\ref{app:3c}) 
with $h_2 \to h_7$.  
\item
$R_{\text{11-2}}=\begin{pmatrix} h_8 & 0 & 0 & 0 \\ 0 & 2h_8 & h_8 & 0 \\ 0 & -h_8 & 0 & 0 \\ h_7 & 0 & 0 & h_8 \end{pmatrix}$ becomes $R_{\text{11-1}}$ by (\ref{app:3a}) 
with $h_7 \to h_2$.  
\item
$R_{\text{11-3}}=\begin{pmatrix} h_8 & 0 & 0 & h_2 \\ 0 & 2h_8 & h_8 & 0 \\ 0 & -h_8 & 0 & 0 \\ 0 & 0 & 0 & h_8 \end{pmatrix}$ becomes $R_{11}$ by (\ref{app:3a}) 
with $h_2 \to h_7$.  
\item
$R_{\text{11-4}}=\begin{pmatrix} h_1 & 0 & 0 & 0 \\ 0 & 0 & -h_1 & 0 \\ 0 & h_1 & 2h_1 & 0 \\ h_7 & 0 & 0 & h_1 \end{pmatrix}$ becomes $R_{\text{11-2}}$ by (\ref{app:3c}) 
with $h_1 \to h_8$.  
\item
$R_{\text{11-5}}=\begin{pmatrix} h_1 & 0 & 0 & h_2 \\ 0 & 0 & -h_1 & 0 \\ 0 & h_1 & 2h_1 & 0 \\ 0 & 0 & 0 & h_1 \end{pmatrix}$ becomes $R_{\text{11-3}}$ by (\ref{app:3c}) 
with $h_1 \to h_8$. 
\item
 $R_{\text{11-6}}=\begin{pmatrix} h_1 & 0 & 0 & 0 \\ 0 & 0 & h_1 & 0 \\ 0 & -h_1 & 2h_1 & 0 \\ h_7 & 0 & 0 & h_1 \end{pmatrix}$ becomes $R_{11}$ by (\ref{app:3c}) 
with $h_1 \to h_8$.  
\item
$R_{\text{11-7}}=\begin{pmatrix} h_1 & 0 & 0 & h_2 \\ 0 & 0 & h_1 & 0 \\ 0 & -h_1 & 2h_1 & 0 \\ 0 & 0 & 0 & h_1 \end{pmatrix}$ becomes $R_{\text{11-1}}$ by (\ref{app:3c}) 
with $h_1 \to h_8$. 
\end{enumerate}
\item
Class 12:
$R_{12}=\begin{pmatrix} h_1 & 0 & 0 & h_2 \\ 0 & \frac{1-\mathrm{i}}{2}h_1 & 0 & 0 \\ 0 & 0 & \frac{1-\mathrm{i}}{2}h_1 & 0 \\ -\frac{\mathrm{i}h_1^2}{2h_2} & 0 & 0 & -\mathrm{i}h_1 \end{pmatrix}$\\
This falls into $R_{H1,1}$ with $p=1$ and $q=\mathrm{i}$ by the transformation (\ref{app:Rchange}): 
$\kappa (Q\otimes Q) R_{H1,1} (Q\otimes Q)^{-1}=R$ 
with $\kappa=\frac{h_1}{2(1+\mathrm{i})}$ and 
$Q=\begin{pmatrix} \sqrt{(1+\mathrm{i})h_2} & 0 \\ 0 & -\sqrt{h_1} \end{pmatrix}$.\\
The other solution 
$R_{\text{12-1}}=\begin{pmatrix} h_1 & 0 & 0 & h_2 \\ 0 & \frac{1+\mathrm{i}}{2}h_1 & 0 & 0 \\ 0 & 0 & \frac{1+\mathrm{i}}{2}h_1 & 0 \\ \frac{\mathrm{i}h_1^2}{2h_2} & 0 & 0 & \mathrm{i}h_1 \end{pmatrix}$ 
becomes $R_{12}$ by (\ref{app:3a}) and (\ref{app:3b}) with $h_1\to -\mathrm{i}h_1$. 
\end{itemize}

\section{Local invariants, link polynomials and entangling power of $R_{H1,3}$ and $R_{H2,3}$}
\label{app:rh23}
 
Among all the two-qubit braid operators in (\ref{app:RH}), the X-type braid operators analyzed in this paper do not fully capture solutions of the form $R_{H1,3}$ and $R_{H2,3}$. 
For completeness we analyze those cases here. 

\paragraph{$R_{H1,3}$}: 
The local invariants presented in Sec. \ref{g2qu} are computed for $R_{H1,3}$ to give $I_1=2k^2$, $I_{2,4}=I_{2,5}=-2k^4$, $I_{2,8}=4k^4$, and $I_{2,9}=I_{2,10}=2k^4$. 
$R_{H1,3}$ has the eigenvalues $k^2$ and $-k^2$ with multiplicities three and one, respectively. 

The enhancement discussed in Sec. \ref{X-link} is possible with $\mu=I-\frac{p+q}{2k}(X+\mathrm{i}Y)$, $x=\pm k^2$ and $y=\pm 1$. 
Link invariants for a two-strand braid word $\xi=\sigma_1^n$ are $L_R(\sigma_1^n)=4$ for $n$ even, and $\pm 2$ for $n$ odd. 
Scaling the enhanced $R$ by $\frac{1}{k^2}$ realizes the Hecke algebra (\ref{hecke}) with $q=1$. 

Following the steps in Sec. \ref{ePower}, $R_{H1,3}$ acts on the product state $\ket{P}=\left(a_1\ket{0}+b_1\ket{1}\right)\otimes\left(a_2\ket{0}+b_2\ket{1}\right)$ to give the local invariant 
$\det t=k^2(p+q)\left(-ka_1b_1b_2^2+ka_2b_1^2b_2+qb_1^2b_2^2\right)$. The entangling power is obtained as $e_P(R_{H1,3})=\frac19|k|^4|p+q|^2\left(|k|^2+|q|^2\right)$. 
The unitary solutions are found at $p=q=0$ and $|k|=1$, and they do not generate entanglement.

\paragraph{$R_{H2,3}$}:
The local invariants for $R_{H2,3}$ are 
$I_1=2k$, $I_{2,4}=I_{2,5}=-2k^2$, $I_{2,8}=4k^2$, and $I_{2,9}=I_{2,10}=2k^2$. 
$R_{H2,3}$ has the eigenvalues $k$ and $-k$ of multiplicities three and one, respectively.  

The enhancement is possible only when $q=-p$ with $\mu=I$, $x=\pm k$ and $y=\pm 1$. 
Link invariants are $L_R(\sigma_1^n) = 4$ for $n$ even and $\pm 2$ for $n$ odd. 
The Skein relation is read from the identity $R_{H2,3}^2-kR_{H2,3}-k^2{\bf 1}+k^3R_{H2,3}^{-1}=0$. 

The entangling power is computed as $e_P(R_{H2,3})=\frac19\left|ks-pq\right|^2$. 
The unitary solutions are found at $p=q=s=0$ and $|k|=1$, in which case the operator is no longer an entangler.


\end{document}